\documentclass[ aip, jcp, amsmath,amssymb,reprint]{revtex4-1}

\usepackage{graphicx}
\usepackage{dcolumn}
\usepackage{bm}
\usepackage[utf8]{inputenc}
\usepackage[T1]{fontenc}
\usepackage{mathptmx}
\usepackage{etoolbox}
\usepackage{xcolor}

\begin{document}

\title{Sequence disorder-induced first order phase transition in confined polyelectrolytes \\ ~}
\author{V. Stepanyan}
\affiliation{Yerevan State University, Yerevan, Armenia}
\author{A. Badasyan}%
\affiliation{University of Nova Gorica, Nova Gorica, Slovenia
}%
\author{V. Morozov}
\affiliation{Yerevan State University, Yerevan, Armenia
}
\author{Y. Mamasakhlisov}
\email{y.mamasakhlisov@ysu.am}
\affiliation{Yerevan State University, Yerevan, Armenia
}
\affiliation{Institute of Applied Problems of Physics, Yerevan, Armenia
}
\author{R. Podgornik}
\email{podgornikrudolf@ucas.ac.cn}
\affiliation{School of Physical Sciences, University of Chinese Academy of Sciences, Beijing, 100049, China}
\affiliation{CAS Key Laboratory of Soft Matter Physics, Institute of Physics, Chinese Academy of Sciences, Beijing, 100190, China}
\affiliation{Kavli Institute for Theoretical Sciences, University of Chinese Academy of Sciences, Beijing, 100049, China}
\affiliation{Wenzhou Institute of the University of Chinese Academy of Sciences, Wenzhou, Zhejiang 325011, China}
\affiliation{Department of Physics, Faculty of Mathematics and Physics, University of Ljubljana, Jadranska 19, 1000 Ljubljana, Slovenia}

\date{\today}
  

\begin{abstract}
We consider a statistical mechanical model of a generic flexible polyelectrolyte, comprised of identically charged monomers with long range electrostatic interactions, and short-range interactions quantified by a disorder field along the polymer contour sequence, 
which is randomly quenched. The free energy and the monomer density profile of the system for no electrolyte screening are calculated in the case of a system composed of two infinite planar bounding surfaces with an intervening oppositely charged polyelectrolyte chain. We show that the effect of the contour sequence disorder, mediated by short-range interactions, leads to an enhanced localization of the polyelectrolyte chain and a first order phase transition at a critical value of the inter-surface spacing.  This phase transition results in an abrupt change of the pressure from negative to positive values, effectively {\em eliminating} polyelectrolyte mediated bridging attraction.  
\end{abstract}

\maketitle

\section{Introduction}
\label{intro}

Interactions involving biologically relevant heteropolymers, such as nucleic acids, polysaccharides and polypeptides \cite{Leckband_Israelachvili_2001}, are generally of two types \cite{rub_col}. The relatively generic and longer-ranged type interactions originate in the electrostatic charges, being at the origin of the polyelectrolyte phenomenology \cite{DOBRYNIN20051049}, and is often invoked as the primary cause of their solution behavior \cite{Muthukumar_2023}. The more specific, shorter ranged interactions \cite{RevModPhys.82.1887}, related to non-universal chemical identity of the various monomer units \cite{elias2012macromolecules}, cannot be quantified by a single parameter, analogous to the electrostatic charge, and are standardly invoked when the observed polyelectrolyte phenomenology cannot be understood solely based on electrostatic interactions \cite{molecules25071661}. 

The origin of the specific short-range interactions can be understood within different theoretical frameworks as stemming from the differences in polarizability of the monomer subunits in the context of van der Waals interactions between polymers \cite{Lubing}, from the differences in the dissociation properties of ionizable moieties within a generalized charge-regulated electrostatics \cite{borkovec2001ionization}, or within a detailed description of water structuring models accounting for the solvent mediated interactions \cite{Blossey_2022}, to list just a few. While each of these various theoretical frameworks of accounting for specific interactions within the context of polymer solution theory leads to differing predictions on the details of these interactions, they all agree in the sense that these interactions are important to understand the behavior of polymers in aqueous solutions.   

The diversity of these specific interactions, originating from different types of monomer units, such as 4 canonical nucleotides in nucleic acids \cite{bloomfield2000nucleic} or 20 canonical amino acids in polypeptides \cite{finkelstein2016protein}, is therefore much more pronounced in terms of short range non-electrostatic interactions than long range electrostatic effects \cite{RevModPhys.82.1887} and is closely related to the sequence of monomers along the chain,  leading to interesting effects for polymer-polymer interaction as well as polymer-substrate interaction.  Gauging the relative importance of the generic long-range Coulomb interaction vs. the sequence specific short-range interaction is particularly important when trying to assess their role in the complicated multi-level problem of RNA folding \cite{fallmann2017recent}, RNA substrate interaction \cite{Guzman}, when formulating realistic models of RNA virus co-assembly \cite{twarock2018modelling}, or manipulating electrostatic repulsion in the background of sequence-specific Watson-Crick base pair interaction of ssRNAs \cite{adhikari2020intra}. In all these cases an interplay of both types of interaction needs to be properly accounted for. 

In what follows we will investigate a model hetero-polyelectrolyte chain, characterized by both generic long-range Coulomb interactions between its equally charged monomers, with superimposed specific short-range interactions depending on the type of the monomer along its sequence. We will assume that the monomer sequence can be modeled as a quenched, random sequence, represented by a single random variable characterized by a Gaussian distribution with zero mean. We will write the free energy of the system within the self-consistent field theory of a confined, single chain within a replica symmetric {\sl Ansatz}, and investigate the interactions between two planar surfaces, charged oppositely from the chain monomers, as a function of the separation between them. We will show that the model confined polymer chain exhibits an additional localization induced by sequence disorder, akin to the Anderson localization, and will assess the relative importance of the disorder effects superimposed on an electrostatic background.

\section{Model and Methods} \label{themodel}

\subsection{Model}

\begin{figure}
\includegraphics[width=.45\textwidth]{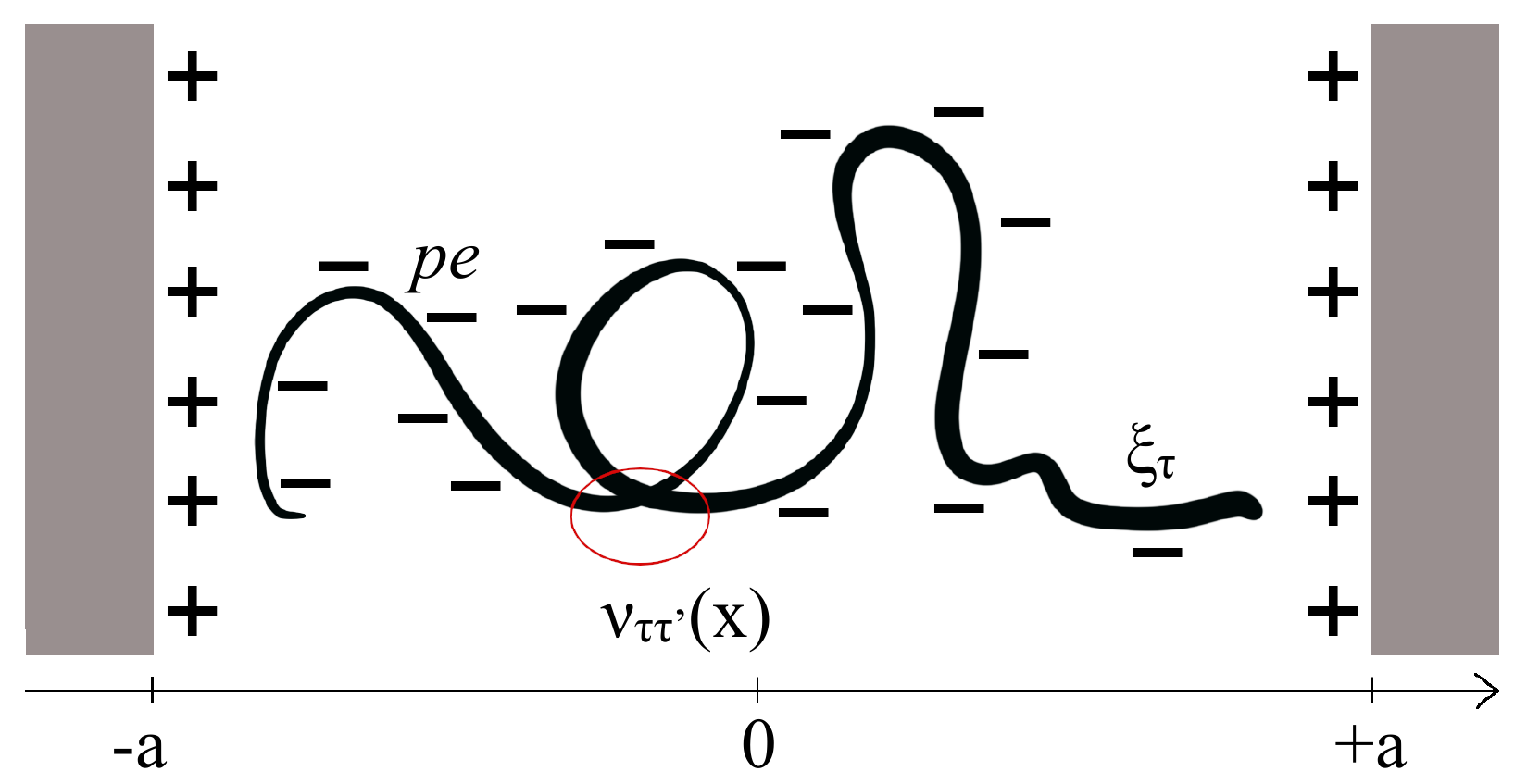}
\caption{\label{polym_in} The negatively charged polyelectrolyte chain is confined between two positively charged infinite plates with surface charge density $\sigma$. The charge per monomer is $pe$, where $e$ is the unit charge, and the short-range interaction between two parts of the chain is $v_{\tau\tau'}({\bf x})=v_0\xi_{\tau}\xi_{\tau'}\delta({\bf x})$, where $\xi_{\tau}$ is the random sequence disorder variable and $\tau$ is the position along the contour of the chain.}%
\end{figure}

We consider a generic flexible hetero-polyelectrolyte, {\it e.g.} ssRNA, see Fig. \ref{polym_in}, confined between two planar surfaces, comprised of equally charged monomers interacting {\sl via} long-range Coulomb interactions, as well as short range "chemical" interactions dependent on the monomer sequence. While the charge of the monomers is assumed to be the same for all the monomers, the short range interactions are sequence specific, with this specificity being encoded in the sequence disorder. This implies that instead of considering a detailed sequence of monomers characterized by different chemical identities, we characterize the strength of the short-range interactions by a random, Gaussian distributed, variable. Similar systems without the sequence disorder affecting the short-range interactions  \cite{borukh, shafir, siber_rudipod08, rudi91,andelman,PhysRevE.107.024503} and the problem of a polymer chain with excluded volume interactions in quenched random media \cite{Muthudisorder,C1PC90001C,bratko1995polyelectrolyte} have been studied before. Our main goal here is to address the interplay between the disordered short-range interactions and long-range Coulomb interactions. In order to simplify the calculations we consider a salt free system with the surfaces oppositely charged from the polymer chain.  

The charge per monomer is assumed to be $pe$, where $e$ is the electron charge and $0 < p \leq 1$. The value $p=1$ corresponds to the fully charged polyelectrolyte, while $p < 1$ describes the weak polyelectrolyte, where ion binding/condensation partially neutralises the charges. The position of a monomer is characterized by a continuous Cartesian coordinate ${\bf r}(\tau)$, where $\tau\in[0,N]$ marks the position along the contour of the chain, and $N$ is the dimensionless contour length of the chain. The random variable characterizing the strength of the short-range interaction is $\xi_{\tau}$. $\{\xi\}$s are considered to be independent, quenched random variables with the probability distribution 
\begin{equation}\label{distribfunc} 
	{\cal P}\{\xi\}=\prod_{\tau}p(\xi_{\tau}),
\end{equation} 
where the sequence disorder is supposed to be Gaussian in order to make the formalism tractable, therefore
\begin{equation} 
	p(\xi_{\tau})=\frac{\mathit{e}^{-\frac{(\xi_{\tau} - \bar{\xi})^2}{2\xi^2}}}{\sqrt{2\pi\xi^2}}.
\end{equation}
For the sake of simplicity we restrict ourselves to the vanishing average value, $\bar{\xi}=0$, case. 

The Hamiltonian of the system in this case has a general Edwardsian form with
\begin{eqnarray}\label{hamiltonian}
	&&\beta{\cal H}[{\bf r}(\tau)] = \frac{3}{2\ell^2}\int_0^N \,d\tau(\partial_{\tau}{\bf r}(\tau))^2+\notag \\ 
	&& + \frac{\beta}{2}\int_0^N \,d\tau\int_0^N \,d\tau'v_{\tau\tau'}({\bf r}(\tau)-{\bf r}(\tau'))+\beta V[\{{\bf r(\tau )}\}],
\end{eqnarray}
where $\ell$ is the Kuhn length, $V[\{{\bf r}(\tau)\}]$ describes all the long-range interactions in the system, {\sl i.e.}, the Coulomb interactions between monomers as well as the Coulomb interactions between monomers and the surface, while 
\begin{equation}\label{vzeroxisquared}
v_{\tau\tau'}({\bf x})=v_0\xi_{\tau}\xi_{\tau'}\delta({\bf x}),
\end{equation}
with a positive interaction constant $v_{0}>0$. Thus, the different monomers will attract, but the similar ones will repel. 

\subsection{Partition function and free energy}

We formulate the partition function of this three-dimensional polymer chain by consistently including the contribution of the monomer sequence disorder and analyzing the combined effect of randomly quenched short-range and homogeneous long-range interactions on the confinement free energy.  The partition function of the model polyelectrolyte chain for every fixed sequence $\{\xi\}$  is given by 
\begin{equation} 
	Z\{\xi\}=\int \,{\cal D}[{\bf r}(\tau )]~e^{-\beta{\cal H}[{\bf r}(\tau)]}.
\end{equation} 
In the limit of a very long chain, $N\rightarrow\infty$, consistent with the ground-state dominance {\sl Ansatz}, the disorder part of the free energy obeys the principle of self-averaging yielding 
\cite{binder_young,parisi2023}: 
\begin{equation}\label{freen_self} 
	{\cal F}=-k_{B}T\langle{\ln Z\{\xi\}}\rangle_{{\cal P}},
\end{equation} 
where $\langle{...}\rangle_{{\cal P}}$ stands for the average with the distribution function (Eq. \ref{distribfunc}). Self-averaging physically means that the distribution of the free energy has a vary narrow peak in the vicinity of its maximum, corresponding to the mean value of the free energy (Eq. \ref{freen_self}). 

The quenched free energy (Eq. \ref{freen_self}) can now be estimated using the 
{\em replica trick} \cite{binder_young,parisi2023}
\begin{equation}\label{-beta_F}
-\beta{\cal F}=\lim_{n\rightarrow 0}\frac{\langle{Z\{\xi\}^n}\rangle_{{\cal P}}-1}{n},
\end{equation}
where $\beta=(k_{B}T)^{-1}$.  In order to calculate the $n$-replica partition function 
\begin{eqnarray}\label{Z^n}
&&\langle{Z\{\xi\}^n}\rangle_{{\cal P}} =\notag \\ 
& & \qquad  = \int \,{\cal D}{\bf r}
e^{-\frac{3}{2\ell^2}\sum_{a=1}^n\int_0^N \,d\tau(\partial_{\tau}{\bf r}^a(\tau))^2- 
\beta\sum_{a=1}^n V\{{\bf r}^a\}}\times \notag \\ 
& & \quad \int \,{\cal D}\xi{\cal P}\{\xi\}e^{-\frac{\beta v_0}{2}\int_0^N \,d\tau\int_0^N \,d\tau'\xi_{\tau}\xi_{\tau'}\sum_{a=1}^n
\delta({\bf r}^a(\tau)-{\bf r}^a(\tau'))},
\end{eqnarray}
\noindent and thus to estimate the free energy, several order parameters need to be introduced, namely, the  inter-replica overlap $q_{ab}({\bf x},{\bf x}')$, with $a<b$, the density of $a$-th replica monomers $\rho_a({\bf x})$, and the density of the random variable $\xi_{\tau}$ of $a$-th replica monomers $m_a({\bf x})$ as
\begin{eqnarray}\label{q_rho_m}
q_{ab}({\bf x},{\bf x}')&=&\int_0^N \,d\tau\delta({\bf x}-{\bf r}^a(\tau))\delta({\bf x}'-{\bf r}^b(\tau))\notag \\ 
\rho_a({\bf x})&=&\int_0^N \,d\tau\delta({\bf x}-{\bf r}^a(\tau))\notag \\
m_a({\bf x})&=&\int_0^N \,d\tau\xi_{\tau}\delta({\bf x}-{\bf r}^a(\tau)).
\end{eqnarray}
Introducing next the Lagrange multipliers ${\hat q}_{ab}({\bf x},{\bf x}')$ and ${\hat \rho}_a({\bf x})$, conjugated to the first two order parameters in (Eq. \ref{q_rho_m}), allows us to rewrite the  $n$-replica partition function as \cite{g.o.p}
\begin{eqnarray}\label{Z^n_ord_params}
\langle{Z\{\xi\}^n}\rangle_{{\cal P}} &&\propto  e^{-\frac{nV}{2\tilde{v}}\ln(2\pi\beta v_0)} \times \nonumber\\
&& \times \int \,{\cal D}{\rho}\,{\cal D}{{\hat\rho}}\,{\cal D}{q}\,{\cal D}{\hat q}
~e^{g(\rho,{\hat \rho},q,{\hat q})+\ln\zeta({\hat \rho},{\hat q})},
\end{eqnarray}
where $\tilde{v}$ is the monomer volume. The explicit expression for the functional $g(\rho,{\hat \rho},q,{\hat  q})+\ln\zeta({\hat \rho},{\hat q})$, as given by (Eq. \ref{g}), allows us to  obtain the final form of the free energy functional as :
\begin{eqnarray}\label{cal_F}
-\beta {\mathcal F}(\rho,{\hat \rho},q,{\hat  q}) = \lim_{n\rightarrow 0}\frac{1}{n}\bigg[-\frac{nV}{2\tilde{v}}\ln(\beta v_0)-\notag\\
\frac{1}{2}{\rm Tr}\ln \hat A - \frac{1}{2}{\rm Tr}\ln\bigl\{I + 
\xi^2\hat A^{-1}\hat B\bigr\} 
-\notag\\
\beta \sum_a W_{el}^a + i\sum_a\langle\rho_a|\hat\rho_a\rangle 
+i\sum_{a<b}\langle q_{ab}|\hat q_{ab}\rangle + \ln\zeta\bigg]
\end{eqnarray}
where the operators $\hat A$ and $\hat B$ have been defined as $n\times n$ operator matrices $\langle{\bf x}|{\hat A}_{ab}|{\bf x}'\rangle =\delta_{ab}\delta({\bf x}-{\bf x}')[\frac{1}{\beta v_0}+\xi^2\rho_a({\bf x})]$ and $\langle{\bf x}|{\hat B}_{ab}|{\bf x'}\rangle= (1-\delta_{ab})q_{ab}({\bf x},{\bf x}')$. To estimate the free energy of the system in the self-consistent field approximation we need to minimize the functional (Eq. \ref{cal_F}) over the fields $\rho$, ${\hat \rho}$, $q$, and ${\hat  q}$. For relevant details and derivations see the Appendix \ref{appendix:FEF}.

The term $\ln\zeta({\hat \rho},{\hat q})$ describes the entropy of the polymeric chain in terms of a  ``quantum-like" particle confined in the restricted volume. The energetic spectrum of such a system is expected to have a gap in the energy spectrum between the ground state and the rest of the spectrum. Thus, the free energy of the system can be calculated using the ground state dominance {\sl Ansatz}, where the partition function (Eq. \ref{Z^n_ord_params}) is dominated by the ground state "wave function" $\psi({\bf r})$ with energy $\mathcal{E}_0$ \cite{g.o.p} which is equivalent to the one-replica density field in the equation (Eq. \ref{edwards0}) (see below)\begin{equation}
\langle{Z\{\xi\}^n}\rangle_{{\cal P}} \propto e^{-N\mathcal{E}_0}\int \,d^3 R \psi({\bf 0})\psi({\bf R}).
\end{equation}
The ground state dominance {\sl Ansatz} is known to work in the case when the polymer length is long enough compared to the separation between the bounding surfaces \cite{Li_2018}. 

\subsection{Replica - symmetric solution}\label{RSS}

In order to estimate the free energy of the system we maximize the functional $g(\rho,{\hat \rho},q,{\hat q})+\ln\zeta({\hat \rho},{\hat 
q})$ from (Eq. \ref{Z^n_ord_params}). We will restrict ourselves to the  replica - symmetric order parameters: $q_{ab}({\bf x},{\bf x}') = 
q({\bf x},{\bf x}')$,  $\hat{q}_{ab}({\bf x},{\bf x}') = \hat{q}({\bf 
x},{\bf x}')$,  $\rho_{a}({\bf x}) = \rho({\bf x})$ and $\hat{\rho}_{a}({\bf x}) = \hat{\rho}({\bf x})$. 
See Appendix Eq. \ref{appendix:RSS} for details.

The electrostatic part of the free energy within the mean-field Poisson-Boltzmann electrostatics can be obtained as \cite{borukh,shafir,siber_rudipod08}
\begin{align}\label{Wel_main}
W_{el}(\rho,\varphi,c^{\pm})=\int\,d^3{\bf x}\bigg\{-\frac{\epsilon\epsilon_0}{2}(\nabla\varphi({\bf x}))^2+\notag \\ 
\varphi({\bf x})\bigg[ec^{+}({\bf x})-ec^{-}({\bf x})
-pe\rho({\bf x})+\rho_{surf}({\bf x})\bigg]+\notag \\ 
\sum_{i=\pm}\bigg[k_BT(c^i({\bf x})\ln c^i({\bf x})-\notag \\ 
c^i({\bf x})-(c_0^i\ln c_0^i-c_0^i))-\mu^i(c^i({\bf x})-c_0^i)\bigg]\bigg\},
\end{align}
where $\varphi({\bf r})$ is electrostatic potential, $c^{\pm}$ are the concentrations of $\pm$ monovalent electrolyte ions in the bathing solution, with $c_0^{\pm}$ being their bulk concentrations, $\mu^{\pm}$ are their chemical potentials, $\varepsilon\varepsilon_0$ is the permittivity of water, and $\rho_{surf}$ is the charge density over the bounding surfaces, confining polyelectrolyte. The interconnection between the electrostatic field and polymer density is provided by term $-pe\varphi({\bf x})\rho({\bf x})$ in (Eq. \ref{Wel_main}). The ground state dominance approximation suppresses polymer density fluctuations and the Poisson-Boltzmann approximation is expected to be reasonable for a monovalent electrolyte. Finite size chain would lead to a decrease of the energy gap between the ground state and first excited state, thus invalidating the ground state dominance approximation \cite{Li_2018}, while the presence of multivalent electrolyte would invalidate the saddle-point approximation and thus the Poisson-Boltzmann approximation \cite{Perspective}.

Solving the system of the saddle - point equations (Eqs. \ref{rho},\ref{varphi},\ref{c}) we get two separate equations for the electrostatic and configurational degrees of freedom, corresponding to the polymer {\em Poisson-Boltzmann equation} \cite{markovich2021charged} for the saddle-point electrostatic potential $\varphi({\bf r})$
 \begin{equation}\label{poissonboltzmann3d}
\varepsilon\varepsilon_0\nabla^2\varphi({\bf r})=2ec_0\sinh(\beta e \varphi({\bf r})) + peN\psi({\bf r})^2 - \rho_s({\bf r})
\end{equation}
as well as the {\em Edwards equation} for the saddle-point polymer one-replica density field $\psi({\bf r})$
\begin{eqnarray}\label{edwards0}
\mathcal{E}_0\psi({\bf r})=\bigg[-\frac{\ell^2}{6}\nabla^2 -\beta pe\varphi({\bf r})+\frac{1}{2}\frac{1}{\mu+N\psi({\bf r})^2}\nonumber\\
\bigg(\frac{1}{\tilde{v}}+\frac{\kappa}{(1-\kappa)^2}\frac{2\mu\psi^2({\bf r})+N\psi({\bf r})^4 }{\mu + N\psi({\bf r})^2}\bigg)\bigg]\psi({\bf r}),
\end{eqnarray}
corresponding to the monomer density $\rho({\bf r})=N\psi({\bf r})^2$ and we introduced te parameter 
$$\kappa=\int d^3x \frac{N\psi({\bf x})^4}{\mu + N\psi({\bf 
x})^2} ~~~~ {\rm and} ~~~~\mu = (\beta v_0 \xi^2)^{-1}.$$Upon substitution of the saddle - point equations 
(see Appendix, Eq.~\ref{rho} - \ref{c}) into (Eq.~\ref{cal_F}), we remain with 
\begin{equation}\begin{split}\label{cal_F_final_3d}
	\mathcal{F} &= \frac{N l^2}{6\beta}\int (\nabla \psi)^2 d^3{\bf x} -\frac{V}{2\beta\tilde{v}}\ln{\mu} + \frac{1}{2\beta} \bigg(\frac{\kappa}{1-\kappa} + \ln{(1-\kappa)}\bigg) +\\&+ \frac{1}{2\beta\tilde{v}}\int d^3{\bf x}\ln{(\mu + N\psi^2({\bf x}))}+\frac{\varepsilon\varepsilon_0}{2}\int d^3{\bf x}(\nabla \varphi({\bf x}))^2,
\end{split}\end{equation}
which is the form of the free energy that we will use in what follows. 

\subsection{No-salt regime and planar confinement geometry} \label{simplest}

Let us now consider the model polyelectrolyte chain confined between two infinite planar surfaces  of area $S$ separated by a separation $2al$, so that $2a$ is the inter -- plate separation in the units of the Kuhn segment length $l$, assuming the limiting salt-free case, corresponding to zero added electrolyte concentration $c_0=0$. This is the simples possible model of a confined polyelectrolyte \cite{rudi91}. We will further assume that the homogeneous surface charge density is described by
\begin{equation}
	\rho_s({\bf r}) = \sigma (\delta(zl - al) + \delta(zl + al)),
\end{equation}
\noindent and that
\begin{eqnarray}
	\varphi(z) &=& \varphi({\bf r}) \\
	\Phi(z) &=& \psi({\bf r})\sqrt{Sl}, 
\end{eqnarray}
\noindent where due to electroneutrality, $\sigma = \frac{\tau N}{2S}$, and $\tau = pe$. We also introduce the dimensionless coordinates  ${\bf r} = (xl, yl, zl)$ and stipulate the polymer density normalization $\int_{-a}^{a} dz ~\Phi^2(z) = 1$. With these definitions the 
Poisson-Boltzmann equation (Eq. \eqref{poissonboltzmann3d}) becomes simply the Poisson equation in the form 
\begin{equation}\label{poissonboltzmann1d}
	\frac{d^2}{dz^2} \varphi(z) = \frac{\sigma 
	l}{\varepsilon\varepsilon_0} 
	\bigg(2\Phi^2(z)-\delta(z-a)-\delta(z+a)\bigg),
\end{equation}
\noindent and thus has an obvious solution 
\begin{equation}\label{poissonboltzmannsol}
	\varphi(z) = \frac{\sigma l}{\varepsilon\varepsilon_0} \bigg[ 
	\int_{-a}^{a}\!\!\!|z-z'| \Phi^2(z') dz' - m(z,a)\bigg].
\end{equation}
with $m(z,a) \equiv \max\big(z-a, -a-z, 0\big)$. In the same limit the Edwards equation (Eq. \eqref{edwards0}) can be cast into the following simplified form
\begin{equation}
\begin{split}
	\mathcal{E}_0 \Phi(z) &= \bigg[-\frac{1}{6}\frac{d^2}{dz^2} - 
	\beta\tau\varphi(z) + \frac{1}{2\tilde{v}}\frac{1}{\mu+\frac{2\sigma}{\tau l}\Phi^2(z)} +\\&+ \frac{\Phi^2(z)}{Sl}\frac{\kappa}{2(1-\kappa)^2}\frac{2\mu+\frac{2\sigma}{\tau l}\Phi^2(z)}{\big(\mu+\frac{2\sigma}{\tau l}\Phi^2(z)\big)^2}\bigg]\Phi(z),
\end{split}
\end{equation}
\noindent where
\begin{equation}
	\kappa = \int_{-a}^{a} dz \frac{2\sigma\Phi^4(z)}{\mu\tau l+2\sigma\Phi^2(z)}.
\end{equation}
As we consider the bounding plates to be of infinite extension, $S\to\infty$, the Edwards equation is further simplified and assumes the form
\begin{equation}\label{edwards1d}
	\frac{1}{6}\Phi''(z) = -\mathcal{E}_0\Phi(z)  - 
	\beta\tau\varphi(z)\Phi(z) + \frac{1}{2\tilde{v}}\frac{\Phi(z)}{\mu+\frac{2\sigma}{\tau l}\Phi^2(z)}.
\end{equation}
Following the linearization procedure for (Eq. \eqref{poissonboltzmannsol}), 
suggested in Ref. \cite{rudi91}, we get for $z\in[0, a]$ an even simpler version of the Edwards equation that can be  written as
\begin{equation}\label{schrodingerlike}
	\frac{1}{6}\Phi''(z) = -\bigg(\mathcal{E}_0  + \frac{\beta\sigma\tau 
	l}{\varepsilon\varepsilon_0}z\bigg)\Phi(z) + \frac{1}{2\tilde{v}}\frac{\Phi(z)}{\mu+\frac{2\sigma}{\tau l}\Phi^2(z)},
\end{equation}
\noindent coupled with normalization
\begin{equation}\begin{split}
2\int_{0}^{a} dz~ \Phi^2(z) &= 1
\end{split}\end{equation}
and the boundary conditions for the ground state "wave function" 
\begin{equation}\begin{split}
	\Phi'(0) = 0 \;\;\; \Phi(a) &= 0.
\end{split}\end{equation}
and the first excited states
\begin{equation}\begin{split}
\Phi(0) = 0 \;\;\; \Phi(a) &= 0.
\end{split}\end{equation}
Furthermore, we introduce the following notations: 
	$\lambda_B = {\beta e^2}/{4\pi \varepsilon_0\varepsilon}, 	\nu = {N}/{(Sl)}$ and $\alpha = 12\pi p^2l^2\lambda_B$,
where $\lambda_B$ is the Bjerrum length and $\nu$ is the  monomer concentration. The final form of the Edwards equation is then given in the form of a non-linear Schr\" odinger equation  	
\begin{equation}\label{schrodingerlike1}
	\Phi''(z) = -\bigg(6\mathcal{E}_0  + \alpha\nu ~z\bigg)\Phi(z) + 
	\frac{3}{\tilde{v}}\frac{\Phi(z)}{\mu+\nu\Phi^2(z)},
\end{equation}
which is the equation that we will solve numerically in what follows.

\subsection{Reduced Free Energy}

Using (Eq. \eqref{cal_F_final_3d}) and the results from the previous section, we then get the surface density of the free energy for the salt-free case in the following form
\begin{equation}\begin{split}
	A&=\frac{\mathcal{F}}{S} = \frac{2\sigma}{\tau\beta}\mathcal{E}_0 - \frac{\sigma}{\tilde{v}\beta\tau}\int_{-a}^{a}dz\frac{\Phi^2(z)}{\mu + \frac{2\sigma}{\tau l}\Phi^2(z)} + \\&\frac{1}{2S\beta}\bigg(\frac{\kappa}{1-\kappa}+\ln{(1-\kappa)}\bigg) +\\&+ \frac{l}{2\tilde{v}\beta}\int_{-a}^{a} dz \ln{\bigg(1 + \frac{2\sigma}{\tau l \mu}\Phi^2(z) \bigg)}+\\&\sigma\bigg(\varphi(a)+\int_{-a}^{a}\Phi^2(z)\varphi(z)dz\bigg).
\end{split}\end{equation}
After taking the $S\to\infty$ limit and using  the notations introduced in (Eq. \ref{schrodingerlike1}) together with $ \tilde{A} = A/k_BT\nu l$, we obtain:
\begin{equation}\begin{split}\label{reducedfreeenergy}
	\tilde{A}&= \mathcal{E}_0 + \frac{\alpha\nu}{12}a + 
	\frac{\alpha\nu}{12}\int_{-a}^adz\int_{-a}^{a}dz'|z-z'|\Phi^2(z')\Phi^2(z) - \\ &- \int_{-a}^{a}dz\frac{\Phi^2(z)}{2\tilde{v}\mu + 2\tilde{v}\nu\Phi^2(z)}+  \frac{1}{2\tilde{v}\nu}\int_{-a}^{a} dz \ln{\big(1 + \frac{\nu}{\mu}\Phi^2(z) \big)}.
\end{split}\end{equation}
No additional analytic approximations are feasible and/or appropriate at this point. The integrals indicated above can only be preformed numerically with the numerical solutions of the non-linear Schr\" odinger equation (Eq. \ref{schrodingerlike1}).

\section{Results}

\begin{figure}
\includegraphics[width=1\linewidth]{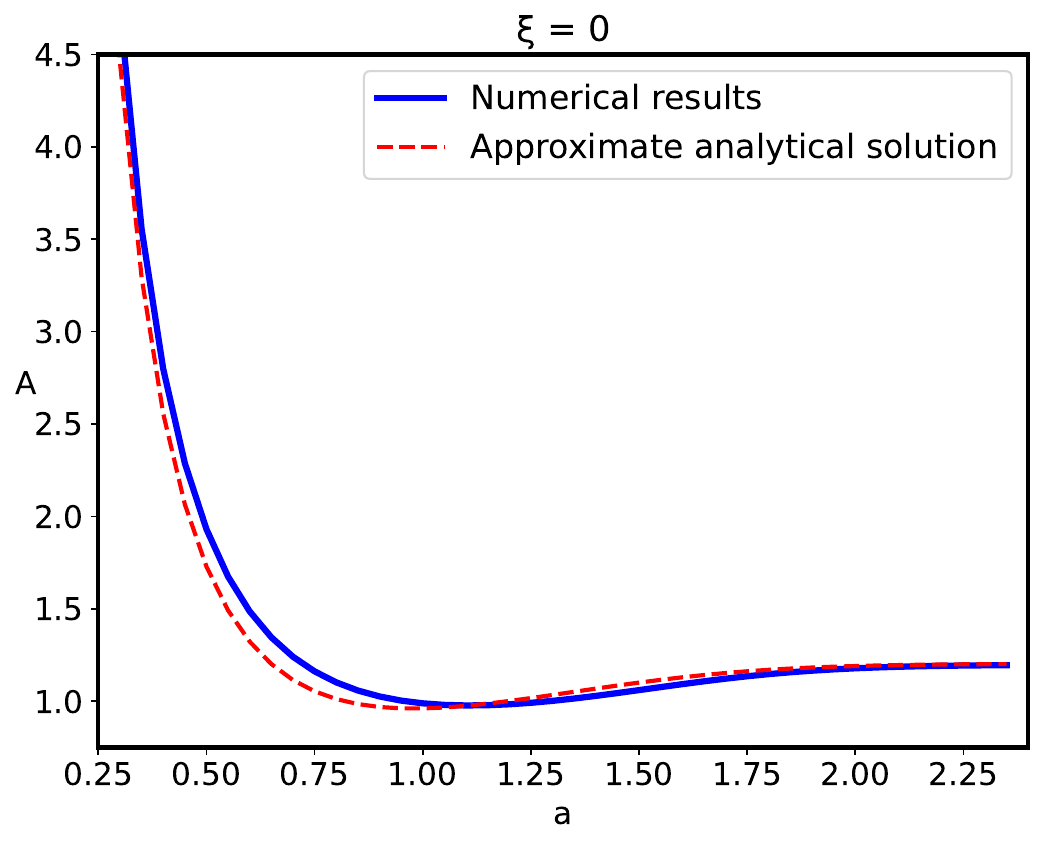}
\includegraphics[width=1\linewidth]{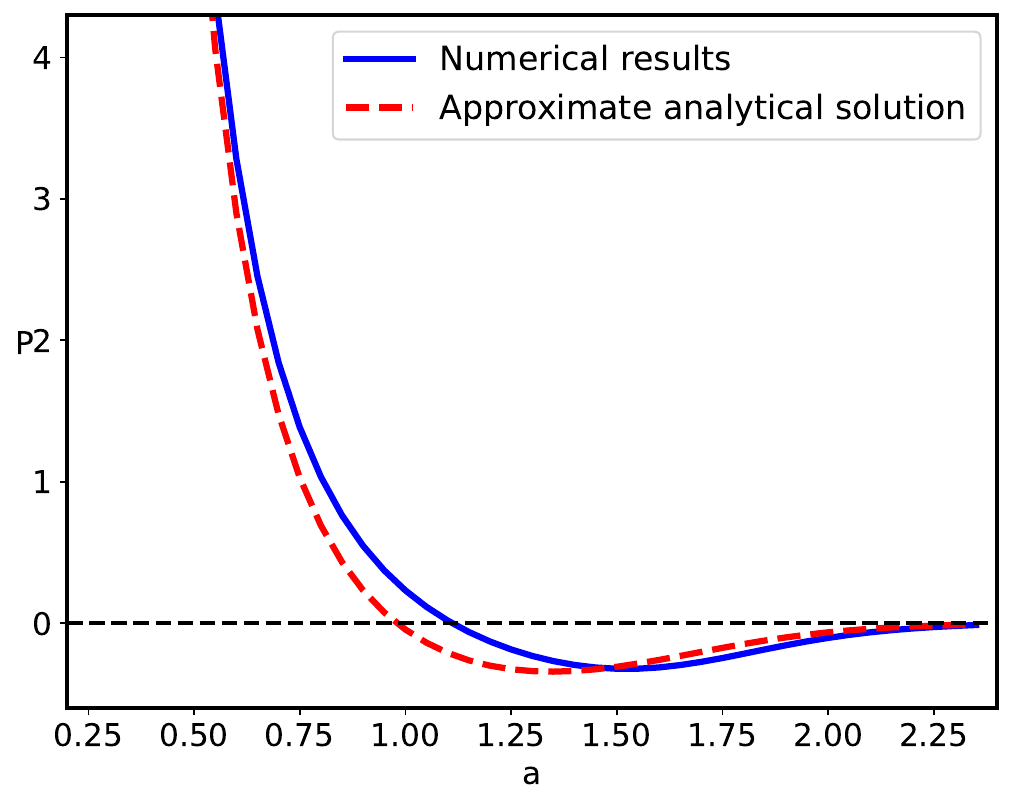}%
\caption{\label{fig_free} Free Energy(upper panel) and pressure (bottom panel) without disorder, in the pure state ($\xi=0.0$) as a function of $a$, one-half of the inter-surface separation $2a$. The blue(solid) lines are calculated numerically based on the Eq.(\ref{reducedfreeenergy}). The red(dashed) lines describe the results from Ref. \cite{rudi91}.}%
\end{figure}

 First, we compare our numerical results with the previously reported analytical results in Ref.~\cite{rudi91}, where the confined polyelectrolyte without short-range disordered interactions was considered and the limits allowing the analytical solution were analyzed. The reduced free energy obtained in Ref.~\cite{rudi91} can be shown to correspond exactly to (Eq. \eqref{reducedfreeenergy}) with $\xi=0$, while additional linearization of the third term in terms of $a$ results in a simple analytical solution~\cite{rudi91}.

We can see from ~Fig.\ref{fig_free} that these additional approximations are not important and the results are effectively the same. The pressure between the two bounding surfaces, {\sl i.e.}, the force per unit area  is calculated from
\begin{equation}
	P = -\frac{1}{l}\frac{\partial \tilde{A}}{\partial a}.
\end{equation}
Using this equation we calculate numerically the pressure of the system that is presented in ~Fig.\ref{fig_free}, and compared to the analytical results of Ref.~\cite{rudi91}.

\begin{figure} 
\includegraphics[width=1\linewidth]{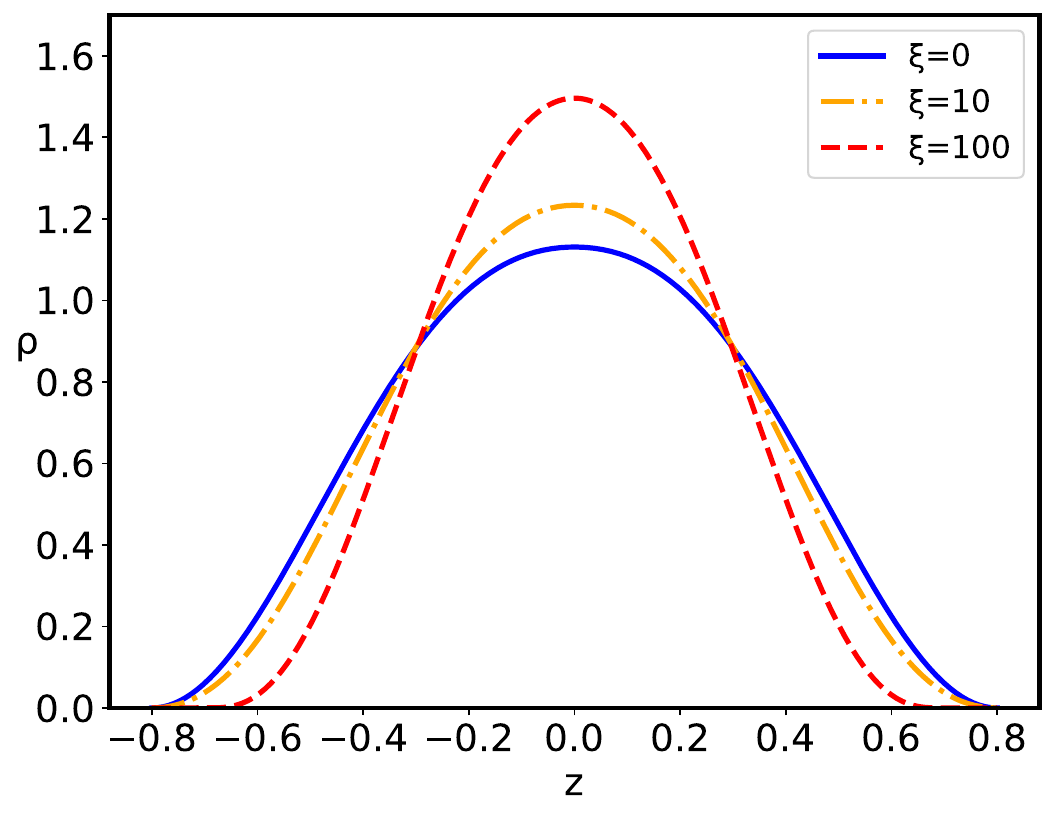}
	\caption{\label{fig_dens_0.8} Dimensionless monomer density profile $\Phi^2(z)$ for $a=0.8$. Blue(solid) line corresponds to the pure system, $\xi = 0$, while the disordered system follows orange(dashed dotted) line - $\xi=10$, and red(dashed) line - $\xi=100$.}
\end{figure}

\begin{figure} 
\includegraphics[width=1\linewidth]{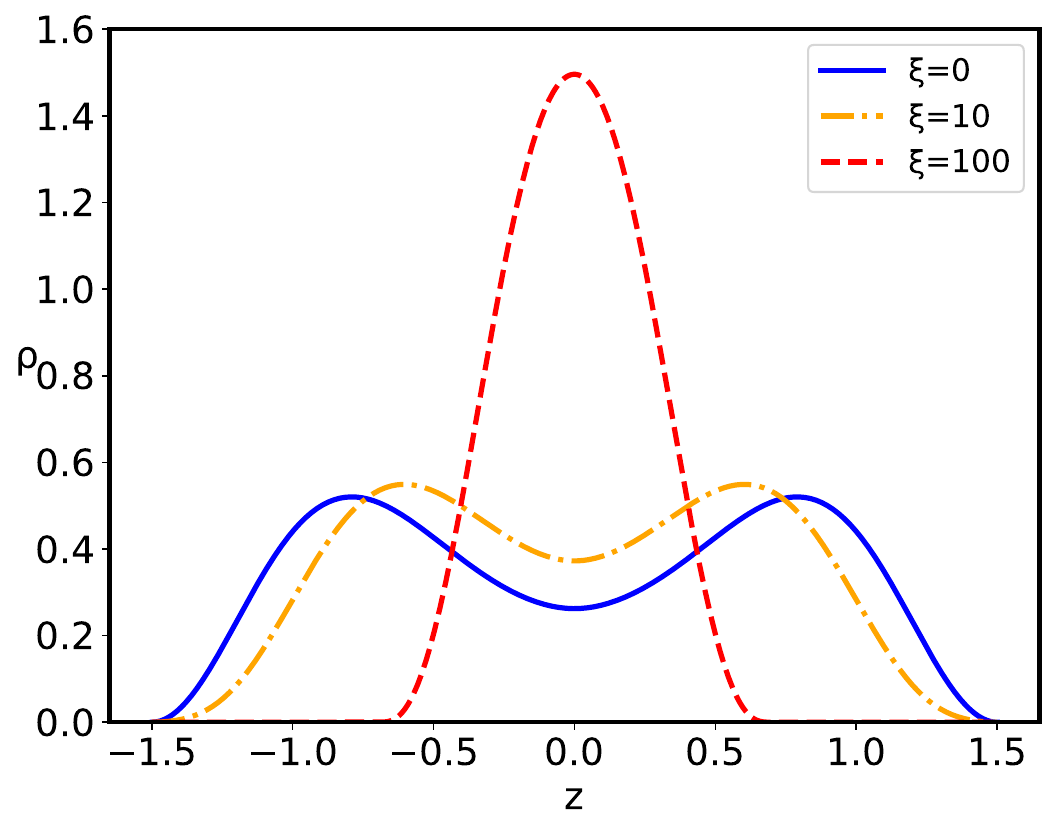}
	\caption{\label{fig_dens_1.25} Dimensionless monomer density profile $\Phi^2(z)$ for $a=1.5$, still below the critical separation $a_c = 1.525$. Blue(solid) line - pure system, orange(dashed dotted) line - disordered one, $\xi=10$, red(dashed) line - disordered one, $\xi=100$.}
\end{figure}

\begin{figure} 
\includegraphics[width=1\linewidth]{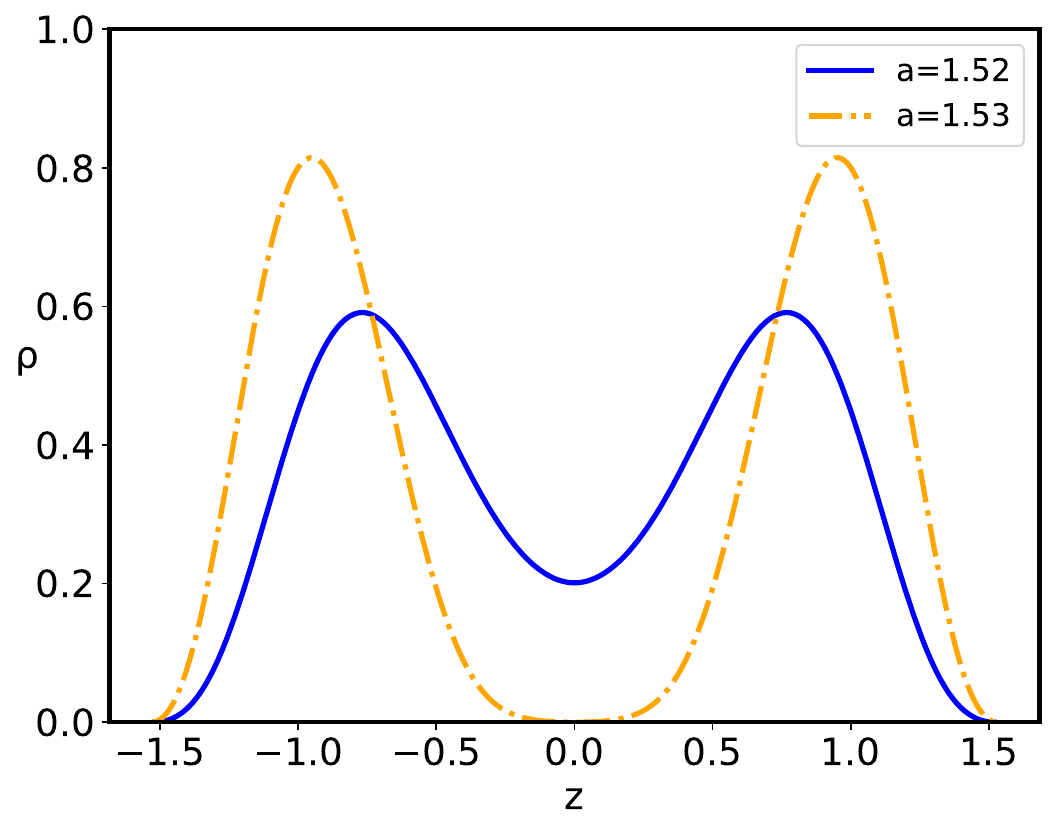}
\caption{Dimensionless monomer density profile below (blue solid line) and above (orange dashed dotted line) the critical inter-surface separation $a_c$. The phase transition of the chain results in a {\em complete depletion} of the monomer density at the midpoint between the bounding surfaces, effectively suppressing the polyelectrolyte bridging induced attraction between the surfaces. }\label{pt_dens}
\end{figure}

The configurational behavior of the confined polyelectrolyte chain with short--range quenched disorder interactions is governed by the interplay between the inter-plate separation $2a$ and the standard deviation of disorder $\xi$. The distribution of the dimensionless polyelectrolyte monomers $\Phi^2(z)$ is presented in ~Figs. \ref{fig_dens_0.8}, \ref{fig_dens_1.25}, and \ref{pt_dens}. For the given value of $\xi$ the behavior of the monomer density distribution on increase of the inter--plate separation is qualitatively the same both for the pure case, corresponding to $\xi=0$, as well as the disordered cases, $\xi\neq 0$, up to the critical inter-plate separation $a_c$: upon further increase of the separation the chain goes through a phase transition from the density profile having two maxima separated by the region of lower non-zero density to the two-maxima density profile, with complete depletion of density at the midpoint between the two bounding surfaces. The critical separation $a_c$ depends on the magnitude of the disorder dispersion $\xi$ as well as on the charge parameters of the system. The main effect of the quenched sequence disorder at the inter-plate separations corresponding to $a < a_c$ is an overall depletion of the monomer density close to the bounding surfaces and accumulation at the relative maximum located at the midpoint between the two bounding surfaces (~Fig. \ref{fig_dens_0.8}) or at the points of the two maxima in case of two-maxima density distribution, leading to an increase of the monomer density in the region of the density maximum (~Fig. \ref{fig_dens_1.25}), i.e., there is an {\em additional localization} of the polyelectrolyte chain on increase of the disorder parameter $\xi^2$. However, for separations corresponding to $a > a_c$, there is also a complete depletion of the monomer at the midpoint between the two bounding surfaces, with preferential accumulation actually at the positions of the relative maxima of the monomer density in the left and right halves of the slit (~Fig. \ref{pt_dens}). 

\begin{figure} 
\includegraphics[width=1\linewidth]{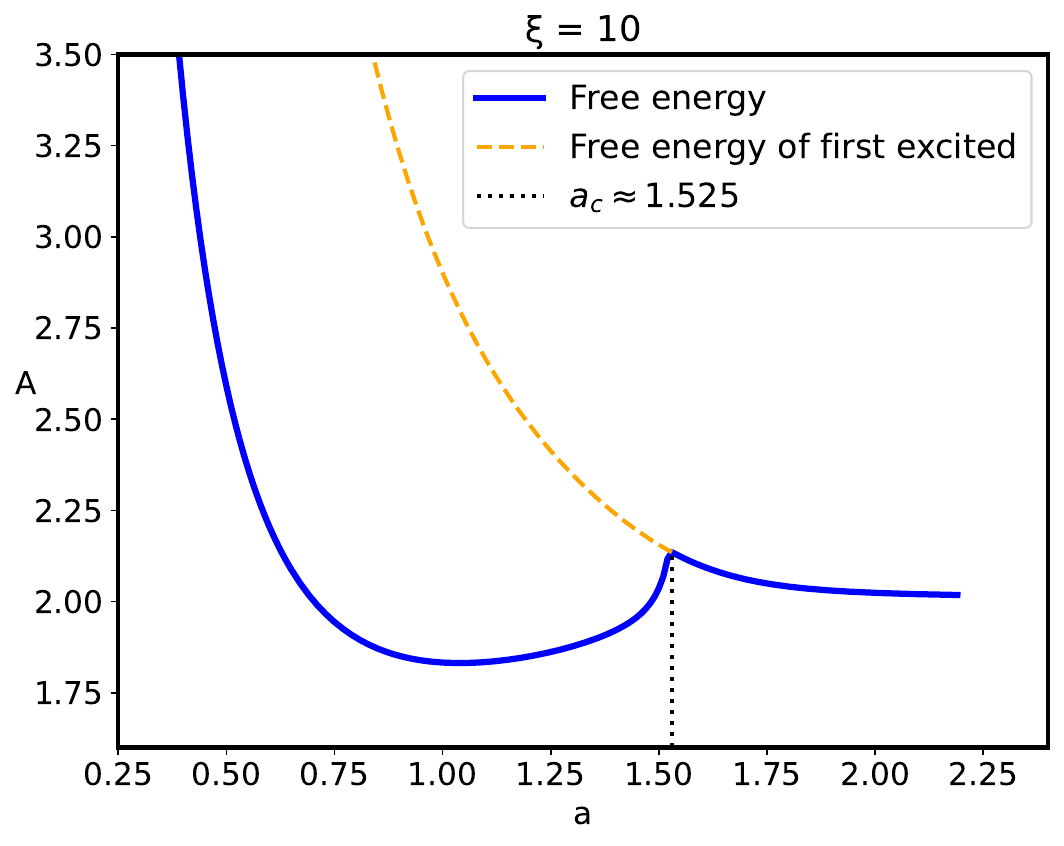}
\caption{\label{fig_free_phase_tr}The free energy dependence on the one-half inter--plate separation $a$ for the ground and first excited states of the equation (\ref{schrodingerlike1}) for the disorder parameter $\xi=10$. Orange dashed line - for the first excited state, blue line - for the ground state. Greater than the critical separation $a_c\approx 1.525$ the original excited state becomes the new ground state.}
\end{figure}

In the presence of disorder the reduced free energy (Eq. \ref{reducedfreeenergy}) is defined by the eigenvalue $\mathcal{E}_0$, corresponding to the "wave function" $\Phi(z)$ of the non-linear Schr\" odinger equation (Eq. \ref{schrodingerlike1}). The free energy behavior corresponding to the ground as well as the first excited states is given in Fig.~\ref{fig_free_phase_tr}, where the singularity of the first derivative of the free energy at $a = a_c$ is clearly shown. Thus, the confined polyelectrolyte with short--range sequence disorder exhibits a phase transition, accompanied by a loss of the original ground state at the critical separation $a_c$. In fact, at the critical separation $a_c$ the ground state free energy actually coincides with that of the first excited state. 

\begin{figure} 
\includegraphics[width=1\linewidth]{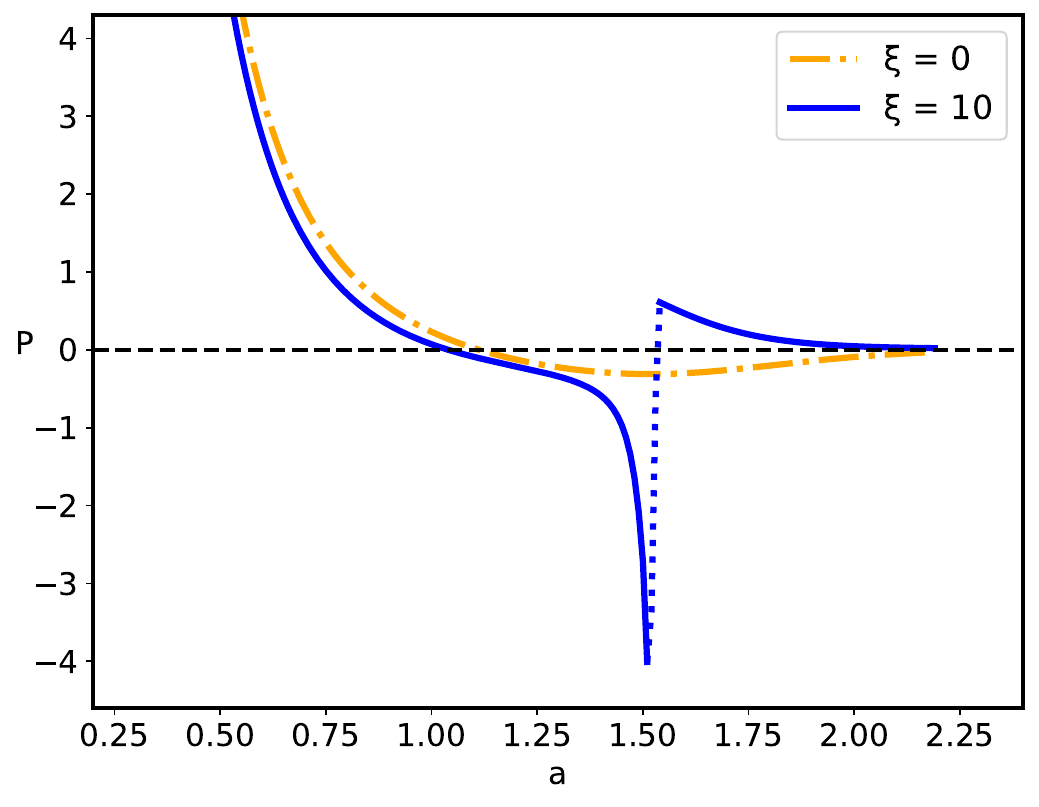}
\caption{Inter-surface pressure dependence on $a$. At the critical separation $a_c = 1.525$ there is a first order-like phase transition of the polyelectrolyte chain characterized by a discontinuous jump in the interaction pressure that changes from attractive to repulsive on the increase of the inter-surface separation beyond the critical value $a_c$. }\label{pt_press}
\end{figure}

The sequence disorder phase transition at a critical inter-surface separation $a = a_c$ is of a first order type, and is accompanied by an abrupt change of the inter-plate pressure (Fig.~\ref{pt_press}), as well as the vanishing of the monomer density at the midpoint between the surfaces (Fig.~\ref{pt_dens}). The pressure behavior for the pure ($\xi = 0$) and disordered ($\xi = 10$) cases (Fig.~\ref{pt_press}) are qualitatively similar below the critical separation $a < a_c$, but diverge sharply for $a > a_c$, when the inter-surface pressure behavior changes drastically from the negative (attractive) pressure in the pure case to the positive (repulsive) pressure in the disordered case. The negative (attractive) pressure for the pure ($\xi = 0$) case is the result of the polyelectrolyte bridging \cite{bridg2006}, where part of the chain is partially adsorbed to one plate and part to the other plate, pulling them together because of the chain connectivity. Thus, the bimodal polyelectrolyte chain density profile between the similarly charged plates (see Fig.~\ref{fig_dens_1.25}) leads to chain-mediated attractions between them. These bridging interactions stem from the fact that the chain is partially adsorbed to both surfaces and the remaining part in between provides an entropic elastic force between them. This situation is very similar to the case of uncharged polymers, except that the polymer-surface interaction is long-range electrostatics for a polyectrolyte chain, and short-range local for an uncharged polymer chain  \cite{DEGENNES1987189}. This furthermore implies that above the critical separation $a_c$ the repulsion between the two positively charged plates is no longer compensated by the polyelectrolyte bridging attraction mediated by the negatively charged polymer chain with the short-range disordered interaction. The reason for this qualitative difference can be traced to a complete charge separation of the polyelectrolyte density between the left and right halves of the inter-surface region, promoted by an increased localization of the monomer density as clearly shown in Fig.~\ref{pt_dens}.

\begin{figure} 
\includegraphics[width=1\linewidth]{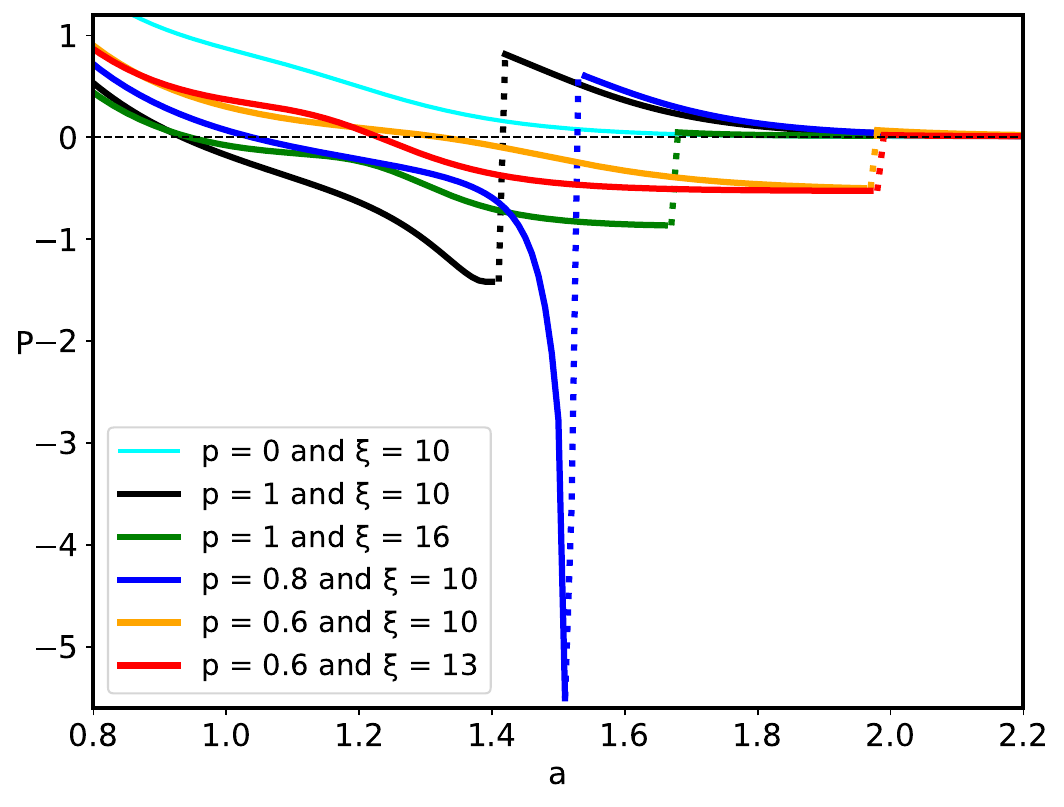}
\caption{Inter-surface pressure dependence on the surface separation $a$ for different values of partial charge $pe$ and different values of the disorder strength $\xi$. Clearly the critical separation $a_c$ of the first order phase transition depends crucially on the values of $p$ and $\xi$, {\sl i.e.}, $a_c = a_c(p, \xi)$ detialed in Table~\ref{tab:cristance}. There is no first-order phase transition for $p=0$.}\label{pt_press_xi}
\end{figure}

The critical separation $a_c$ and the very existence of the phase transition strictly depends on the partial charge $pe$ and, consequently on the surface charge density $\sigma$ as summarized in Fig.~\ref{pt_press_xi}.  A decrease in the partial charge $pe$  for the given value of the strength of disorder $\xi$ leads to an increase in the critical separation $a_c$ (see in Fig.~\ref{pt_press_xi}). Without electrostatics ($p=0$), the inter-plate pressure is always positive (repulsive) without the first-order phase transition. On the other hand, the effect of the strength of disorder $\xi$ on the phase behavior of the system is not unique. We have no phase transition for small enough values of $\xi$, while for higher values of $\xi$, as well as dependent on the partial charge $pe$, the system exhibits a first-order phase transition. In general, the increase in $\xi$, for a given value of partial charge $pe$, leads to an increase in the critical separation $a_c$ (see in Fig.~\ref{pt_press_xi}). The values of the critical separation $a_c$ presented in Fig.~ \ref{pt_press_xi} for the different values of $\xi$ and $pe$ parameters are summarized in Table~\ref{tab:cristance}.

\begin{table}
    \centering
\caption{The dependence of the critical separation $a_c = a_c(p, \xi)$  on parameters $\xi$ and $p$}
\label{tab:cristance}
    \begin{tabular}{|>{\centering\arraybackslash}p{0.3\linewidth}|>{\centering\arraybackslash}p{0.3\linewidth}|>{\centering\arraybackslash}p{0.3\linewidth}|} \hline 
         $p$&  $\xi$& $a_c$\\ \hline 
         1&  10& 1.410\\ \hline 
         1&  16& 1.670\\ \hline 
         0.8&  10& 1.525\\ \hline 
         0.6&  10& 1.970\\ \hline 
         0.6&  13& 1.975\\ \hline
    \end{tabular}
    \end{table}

\section{Discussion}

We developed a self-consistent field (SCF) formalism describing behavior of the polyelectrolyte chain confined between two charged impenetrable bounding surfaces. There is no added salt present in the system and the charges on the polyelectrolyte chain are compensated solely by the charges on the surfaces. The identity of the monomers along the polyelectrolyte chain is described by a random variable, $\xi_{\tau}$, characterized by a Gaussian  distribution of zero mean and variance $\xi^2$. The quenched distribution of monomers along the contour only affects the short range interactions, while the long-range interactions are Coulombic, corresponding to a fixed charge per monomer. 
  
To calculate the free energy and density profile of the monomers, we have used the replica approach and obtained a phase transition of the first order at a critical separation $a_c$ for the replica-symmetric solution. We have considered the polyelectrolyte chain in no-added electrolyte regime, confined between two oppositely charged, impenetrable planar surfaces with the specific aim to asses the contribution of the quenched sequence disorder and the short-range interactions between the monomers.  The obtained results can be interpreted in terms of a  ``quantum-like" Hamiltonian ${\cal {\hat H}}_n=-\frac{\ell^2}{6}\sum_{a}\nabla_{a}^2+\imath\sum_{a}{\hat \rho}_a({\bf x}_a)+\imath\sum_{a<b}{\hat q}_{ab}({\bf x}_a,{\bf x}_b)$, describing an effective particle in the random external field $\imath\sum_{a}{\hat \rho}_a({\bf x}_a)+\imath\sum_{a<b}{\hat q}_{ab}({\bf x}_a,{\bf x}_b)$. The confinement of the polyelectrolyte chain inside the slit with charged walls would thus correspond to the quantum-mechanical particle localized in a potential well. The behavior of the polymer density $\Phi(z)^2$ presented in Figs.(\ref{fig_dens_0.8}, Eq. \ref{fig_dens_1.25}) exhibits the additional localization induced by sequence disorder and consequently by the random effective field in the ``quantum-like" Hamiltonian  ${\cal {\hat H}}_n$, akin to the Anderson localization of the polymer chain in a random medium, as was {\it e.g.} demonstrated in Ref. \cite{shiferaw}. The essence of Anderson localization phenomenon \cite{anderson1958, crisanti_matrix} consists of the fact that under certain conditions the wave function of an electron in a random potential does not spread over the whole space, but is localized in a finite region. In the self-consistent field approximation the confined polyelectrolyte is already localized, but the disorder in the sequence creates an effective random potential that leads to an  additional, disorder driven localization of the "wave function" $\Phi(z)$. 

Besides the additional, disorder driven localization, the sequence disorder also results in a phase transition of a first order type, accompanied by an abrupt change of the inter-surface pressure and monomer density (Fig.~\ref{pt_dens}), on increase of the inter-surface separation beyond the critical separation $a_c$. The pressure behavior for the pure and disordered cases are actually qualitatively similar below the critical separations, $a < a_c$, but at separations $a > a_c$, the inter-surface pressure behavior changes drastically, flipping from a negative pressure in the pure case to a positive one in the disordered case. Above the critical separation $a_c$ the repulsion between the two positively charged plates ceases to be mediated by the polyelectrolyte chain bridging mechanism, due to the complete separation of the polyelectrolyte density between the left and right halves of the slit as a consequence of the disorder-driven localization of the monomer density.  

The work presented above introduces an interesting and important new feature in the phenomenology of polyelectrolyte-driven interactions in charged systems, where the polymer chains exhibit a  "chemical" sequence disorder coupled to short-range interactions.

\acknowledgments
The SCS of Armenia supported this work, grant No. 22AA-1C023 (V.S., Y.M.), grant No. 21AG-1C038 (V.S.), grant No. 21T-1F307 (Y.M.). A.B. and Y.M. acknowledge the partial financial support from Erasmus+ Project No. (2023-1-SI01-KA171-HED-000138423). R.P. acknowledges funding from the Key Project No. 12034019 of the National Natural Science Foundation of China.

\section*{author declarations}
\subsection*{Conflict of Interest}
The authors have no conflicts to disclose.

\subsection*{Author Contribution}

{\bf V. Stepanyan}: Data curation (equal); Formal analysis (equal);  Investigation (equal); Methodology (equal); Soft-
ware (equal); Validation (equal); Visualization (equal).

{\bf A. Badasyan}: Conceptualization (equal); Formal analysis (equal); Methodology (equal); Validation (equal); Writing – original draft (equal).

{\bf V. Morozov}: Conceptualization (equal); Formal analysis (equal); Methodology (equal); Supervision (equal).

{\bf Y. Mamasakhlisov}: Conceptualization (equal); Formal analysis (equal); Investigation
(equal); Methodology (equal); Supervision (equal); Writing – original draft (equal)

{\bf R. Podgornik}: Conceptualization (equal); Formal analysis (equal); Methodology (equal); Supervision (equal); Validation (equal); Writing – original draft (equal).

\section*{Data Availability Statement}
The data that support the findings of this study are available within the article. 

\bibliography{elstat}

\begin{thebibliography}{36}%
\makeatletter
\providecommand \@ifxundefined [1]{%
 \@ifx{#1\undefined}
}%
\providecommand \@ifnum [1]{%
 \ifnum #1\expandafter \@firstoftwo
 \else \expandafter \@secondoftwo
 \fi
}%
\providecommand \@ifx [1]{%
 \ifx #1\expandafter \@firstoftwo
 \else \expandafter \@secondoftwo
 \fi
}%
\providecommand \natexlab [1]{#1}%
\providecommand \enquote  [1]{``#1''}%
\providecommand \bibnamefont  [1]{#1}%
\providecommand \bibfnamefont [1]{#1}%
\providecommand \citenamefont [1]{#1}%
\providecommand \href@noop [0]{\@secondoftwo}%
\providecommand \href [0]{\begingroup \@sanitize@url \@href}%
\providecommand \@href[1]{\@@startlink{#1}\@@href}%
\providecommand \@@href[1]{\endgroup#1\@@endlink}%
\providecommand \@sanitize@url [0]{\catcode `\\12\catcode `\$12\catcode `\&12\catcode `\#12\catcode `\^12\catcode `\_12\catcode `\%12\relax}%
\providecommand \@@startlink[1]{}%
\providecommand \@@endlink[0]{}%
\providecommand \url  [0]{\begingroup\@sanitize@url \@url }%
\providecommand \@url [1]{\endgroup\@href {#1}{\urlprefix }}%
\providecommand \urlprefix  [0]{URL }%
\providecommand \Eprint [0]{\href }%
\providecommand \doibase [0]{http://dx.doi.org/}%
\providecommand \selectlanguage [0]{\@gobble}%
\providecommand \bibinfo  [0]{\@secondoftwo}%
\providecommand \bibfield  [0]{\@secondoftwo}%
\providecommand \translation [1]{[#1]}%
\providecommand \BibitemOpen [0]{}%
\providecommand \bibitemStop [0]{}%
\providecommand \bibitemNoStop [0]{.\EOS\space}%
\providecommand \EOS [0]{\spacefactor3000\relax}%
\providecommand \BibitemShut  [1]{\csname bibitem#1\endcsname}%
\let\auto@bib@innerbib\@empty
\bibitem [{\citenamefont {Leckband}\ and\ \citenamefont {Israelachvili}(2001)}]{Leckband_Israelachvili_2001}%
  \BibitemOpen
  \bibfield  {author} {\bibinfo {author} {\bibfnamefont {D.}~\bibnamefont {Leckband}}\ and\ \bibinfo {author} {\bibfnamefont {J.}~\bibnamefont {Israelachvili}},\ }\href {\doibase 10.1017/S0033583501003687} {\bibfield  {journal} {\bibinfo  {journal} {Quarterly Reviews of Biophysics}\ }\textbf {\bibinfo {volume} {34}},\ \bibinfo {pages} {105} (\bibinfo {year} {2001})}\BibitemShut {NoStop}%
\bibitem [{\citenamefont {Rubinstein}\ and\ \citenamefont {Colby}(2003)}]{rub_col}%
  \BibitemOpen
  \bibfield  {author} {\bibinfo {author} {\bibfnamefont {M.}~\bibnamefont {Rubinstein}}\ and\ \bibinfo {author} {\bibfnamefont {R.~H.}\ \bibnamefont {Colby}},\ }\href@noop {} {\emph {\bibinfo {title} {Polymer Physics}}}\ (\bibinfo  {publisher} {Oxford University Press},\ \bibinfo {address} {New York},\ \bibinfo {year} {2003})\BibitemShut {NoStop}%
\bibitem [{\citenamefont {Dobrynin}\ and\ \citenamefont {Rubinstein}(2005)}]{DOBRYNIN20051049}%
  \BibitemOpen
  \bibfield  {author} {\bibinfo {author} {\bibfnamefont {A.~V.}\ \bibnamefont {Dobrynin}}\ and\ \bibinfo {author} {\bibfnamefont {M.}~\bibnamefont {Rubinstein}},\ }\href {\doibase https://doi.org/10.1016/j.progpolymsci.2005.07.006} {\bibfield  {journal} {\bibinfo  {journal} {Progress in Polymer Science}\ }\textbf {\bibinfo {volume} {30}},\ \bibinfo {pages} {1049} (\bibinfo {year} {2005})}\BibitemShut {NoStop}%
\bibitem [{\citenamefont {Muthukumar}(2023)}]{Muthukumar_2023}%
  \BibitemOpen
  \bibfield  {author} {\bibinfo {author} {\bibfnamefont {M.}~\bibnamefont {Muthukumar}},\ }\href@noop {} {\emph {\bibinfo {title} {Physics of Charged Macromolecules: Synthetic and Biological Systems}}}\ (\bibinfo  {publisher} {Cambridge University Press},\ \bibinfo {year} {2023})\BibitemShut {NoStop}%
\bibitem [{\citenamefont {French}\ \emph {et~al.}(2010)\citenamefont {French}, \citenamefont {Parsegian}, \citenamefont {Podgornik}, \citenamefont {Rajter}, \citenamefont {Jagota}, \citenamefont {Luo}, \citenamefont {Asthagiri}, \citenamefont {Chaudhury}, \citenamefont {Chiang}, \citenamefont {Granick}, \citenamefont {Kalinin}, \citenamefont {Kardar}, \citenamefont {Kjellander}, \citenamefont {Langreth}, \citenamefont {Lewis}, \citenamefont {Lustig}, \citenamefont {Wesolowski}, \citenamefont {Wettlaufer}, \citenamefont {Ching}, \citenamefont {Finnis}, \citenamefont {Houlihan}, \citenamefont {von Lilienfeld}, \citenamefont {van Oss},\ and\ \citenamefont {Zemb}}]{RevModPhys.82.1887}%
  \BibitemOpen
  \bibfield  {author} {\bibinfo {author} {\bibfnamefont {R.~H.}\ \bibnamefont {French}}, \bibinfo {author} {\bibfnamefont {V.~A.}\ \bibnamefont {Parsegian}}, \bibinfo {author} {\bibfnamefont {R.}~\bibnamefont {Podgornik}}, \bibinfo {author} {\bibfnamefont {R.~F.}\ \bibnamefont {Rajter}}, \bibinfo {author} {\bibfnamefont {A.}~\bibnamefont {Jagota}}, \bibinfo {author} {\bibfnamefont {J.}~\bibnamefont {Luo}}, \bibinfo {author} {\bibfnamefont {D.}~\bibnamefont {Asthagiri}}, \bibinfo {author} {\bibfnamefont {M.~K.}\ \bibnamefont {Chaudhury}}, \bibinfo {author} {\bibfnamefont {Y.-m.}\ \bibnamefont {Chiang}}, \bibinfo {author} {\bibfnamefont {S.}~\bibnamefont {Granick}}, \bibinfo {author} {\bibfnamefont {S.}~\bibnamefont {Kalinin}}, \bibinfo {author} {\bibfnamefont {M.}~\bibnamefont {Kardar}}, \bibinfo {author} {\bibfnamefont {R.}~\bibnamefont {Kjellander}}, \bibinfo {author} {\bibfnamefont {D.~C.}\ \bibnamefont {Langreth}}, \bibinfo {author} {\bibfnamefont {J.}~\bibnamefont {Lewis}}, \bibinfo {author}
  {\bibfnamefont {S.}~\bibnamefont {Lustig}}, \bibinfo {author} {\bibfnamefont {D.}~\bibnamefont {Wesolowski}}, \bibinfo {author} {\bibfnamefont {J.~S.}\ \bibnamefont {Wettlaufer}}, \bibinfo {author} {\bibfnamefont {W.-Y.}\ \bibnamefont {Ching}}, \bibinfo {author} {\bibfnamefont {M.}~\bibnamefont {Finnis}}, \bibinfo {author} {\bibfnamefont {F.}~\bibnamefont {Houlihan}}, \bibinfo {author} {\bibfnamefont {O.~A.}\ \bibnamefont {von Lilienfeld}}, \bibinfo {author} {\bibfnamefont {C.~J.}\ \bibnamefont {van Oss}}, \ and\ \bibinfo {author} {\bibfnamefont {T.}~\bibnamefont {Zemb}},\ }\href {\doibase 10.1103/RevModPhys.82.1887} {\bibfield  {journal} {\bibinfo  {journal} {Rev. Mod. Phys.}\ }\textbf {\bibinfo {volume} {82}},\ \bibinfo {pages} {1887} (\bibinfo {year} {2010})}\BibitemShut {NoStop}%
\bibitem [{\citenamefont {Elias}(2012)}]{elias2012macromolecules}%
  \BibitemOpen
  \bibfield  {author} {\bibinfo {author} {\bibfnamefont {H.-G.}\ \bibnamefont {Elias}},\ }\href@noop {} {\emph {\bibinfo {title} {Macromolecules{\textperiodcentered} 1: Volume 1: Structure and Properties}}}\ (\bibinfo  {publisher} {Springer Science \& Business Media},\ \bibinfo {year} {2012})\BibitemShut {NoStop}%
\bibitem [{\citenamefont {Smiatek}(2020)}]{molecules25071661}%
  \BibitemOpen
  \bibfield  {author} {\bibinfo {author} {\bibfnamefont {J.}~\bibnamefont {Smiatek}},\ }\href {\doibase 10.3390/molecules25071661} {\bibfield  {journal} {\bibinfo  {journal} {Molecules}\ }\textbf {\bibinfo {volume} {25}} (\bibinfo {year} {2020}),\ 10.3390/molecules25071661}\BibitemShut {NoStop}%
\bibitem [{\citenamefont {Lu}\ \emph {et~al.}(2015)\citenamefont {Lu}, \citenamefont {Naji},\ and\ \citenamefont {Podgornik}}]{Lubing}%
  \BibitemOpen
  \bibfield  {author} {\bibinfo {author} {\bibfnamefont {B.-S.}\ \bibnamefont {Lu}}, \bibinfo {author} {\bibfnamefont {A.}~\bibnamefont {Naji}}, \ and\ \bibinfo {author} {\bibfnamefont {R.}~\bibnamefont {Podgornik}},\ }\href {\doibase 10.1063/1.4921892} {\bibfield  {journal} {\bibinfo  {journal} {The Journal of Chemical Physics}\ }\textbf {\bibinfo {volume} {142}},\ \bibinfo {pages} {214904} (\bibinfo {year} {2015})},\ \Eprint {http://arxiv.org/abs/https://pubs.aip.org/aip/jcp/article-pdf/doi/10.1063/1.4921892/13254056/214904\_1\_online.pdf} {https://pubs.aip.org/aip/jcp/article-pdf/doi/10.1063/1.4921892/13254056/214904\_1\_online.pdf} \BibitemShut {NoStop}%
\bibitem [{\citenamefont {Borkovec}\ \emph {et~al.}(2001)\citenamefont {Borkovec}, \citenamefont {J{\"o}nsson},\ and\ \citenamefont {Koper}}]{borkovec2001ionization}%
  \BibitemOpen
  \bibfield  {author} {\bibinfo {author} {\bibfnamefont {M.}~\bibnamefont {Borkovec}}, \bibinfo {author} {\bibfnamefont {B.}~\bibnamefont {J{\"o}nsson}}, \ and\ \bibinfo {author} {\bibfnamefont {G.~J.}\ \bibnamefont {Koper}},\ }\href@noop {} {\emph {\bibinfo {title} {Ionization processes and proton binding in polyprotic systems: Small molecules, proteins, interfaces, and polyelectrolytes}}}\ (\bibinfo  {publisher} {Springer},\ \bibinfo {year} {2001})\BibitemShut {NoStop}%
\bibitem [{\citenamefont {Blossey}\ and\ \citenamefont {Podgornik}(2022)}]{Blossey_2022}%
  \BibitemOpen
  \bibfield  {author} {\bibinfo {author} {\bibfnamefont {R.}~\bibnamefont {Blossey}}\ and\ \bibinfo {author} {\bibfnamefont {R.}~\bibnamefont {Podgornik}},\ }\href {\doibase 10.1209/0295-5075/ac7d0a} {\bibfield  {journal} {\bibinfo  {journal} {Europhysics Letters}\ }\textbf {\bibinfo {volume} {139}},\ \bibinfo {pages} {27002} (\bibinfo {year} {2022})}\BibitemShut {NoStop}%
\bibitem [{\citenamefont {Bloomfield}\ \emph {et~al.}(2000)\citenamefont {Bloomfield}, \citenamefont {Crothers},\ and\ \citenamefont {Tinoco}}]{bloomfield2000nucleic}%
  \BibitemOpen
  \bibfield  {author} {\bibinfo {author} {\bibfnamefont {V.~A.}\ \bibnamefont {Bloomfield}}, \bibinfo {author} {\bibfnamefont {D.~M.}\ \bibnamefont {Crothers}}, \ and\ \bibinfo {author} {\bibfnamefont {I.}~\bibnamefont {Tinoco}},\ }\href@noop {} {\emph {\bibinfo {title} {Nucleic acids: structures, properties, and functions}}}\ (\bibinfo  {publisher} {University Science Books},\ \bibinfo {year} {2000})\BibitemShut {NoStop}%
\bibitem [{\citenamefont {Finkelstein}\ and\ \citenamefont {Ptitsyn}(2016)}]{finkelstein2016protein}%
  \BibitemOpen
  \bibfield  {author} {\bibinfo {author} {\bibfnamefont {A.~V.}\ \bibnamefont {Finkelstein}}\ and\ \bibinfo {author} {\bibfnamefont {O.}~\bibnamefont {Ptitsyn}},\ }\href@noop {} {\emph {\bibinfo {title} {Protein physics: a course of lectures}}}\ (\bibinfo  {publisher} {Elsevier},\ \bibinfo {year} {2016})\BibitemShut {NoStop}%
\bibitem [{\citenamefont {Fallmann}\ \emph {et~al.}(2017)\citenamefont {Fallmann}, \citenamefont {Will}, \citenamefont {Engelhardt}, \citenamefont {Gr{\"u}ning}, \citenamefont {Backofen},\ and\ \citenamefont {Stadler}}]{fallmann2017recent}%
  \BibitemOpen
  \bibfield  {author} {\bibinfo {author} {\bibfnamefont {J.}~\bibnamefont {Fallmann}}, \bibinfo {author} {\bibfnamefont {S.}~\bibnamefont {Will}}, \bibinfo {author} {\bibfnamefont {J.}~\bibnamefont {Engelhardt}}, \bibinfo {author} {\bibfnamefont {B.}~\bibnamefont {Gr{\"u}ning}}, \bibinfo {author} {\bibfnamefont {R.}~\bibnamefont {Backofen}}, \ and\ \bibinfo {author} {\bibfnamefont {P.~F.}\ \bibnamefont {Stadler}},\ }\href@noop {} {\bibfield  {journal} {\bibinfo  {journal} {Journal of biotechnology}\ }\textbf {\bibinfo {volume} {261}},\ \bibinfo {pages} {97} (\bibinfo {year} {2017})}\BibitemShut {NoStop}%
\bibitem [{\citenamefont {Poblete}\ \emph {et~al.}(2021)\citenamefont {Poblete}, \citenamefont {An\v{z}e}, \citenamefont {Kandu\v{c}}, \citenamefont {Podgornik},\ and\ \citenamefont {Guzman}}]{Guzman}%
  \BibitemOpen
  \bibfield  {author} {\bibinfo {author} {\bibfnamefont {S.}~\bibnamefont {Poblete}}, \bibinfo {author} {\bibfnamefont {B.}~\bibnamefont {An\v{z}e}}, \bibinfo {author} {\bibfnamefont {M.}~\bibnamefont {Kandu\v{c}}}, \bibinfo {author} {\bibfnamefont {R.}~\bibnamefont {Podgornik}}, \ and\ \bibinfo {author} {\bibfnamefont {H.~V.}\ \bibnamefont {Guzman}},\ }\href {\doibase 10.1021/acsomega.1c04774} {\bibfield  {journal} {\bibinfo  {journal} {ACS Omega}\ }\textbf {\bibinfo {volume} {6}},\ \bibinfo {pages} {32823} (\bibinfo {year} {2021})}\BibitemShut {NoStop}%
\bibitem [{\citenamefont {Twarock}\ \emph {et~al.}(2018)\citenamefont {Twarock}, \citenamefont {Bingham}, \citenamefont {Dykeman},\ and\ \citenamefont {Stockley}}]{twarock2018modelling}%
  \BibitemOpen
  \bibfield  {author} {\bibinfo {author} {\bibfnamefont {R.}~\bibnamefont {Twarock}}, \bibinfo {author} {\bibfnamefont {R.~J.}\ \bibnamefont {Bingham}}, \bibinfo {author} {\bibfnamefont {E.~C.}\ \bibnamefont {Dykeman}}, \ and\ \bibinfo {author} {\bibfnamefont {P.~G.}\ \bibnamefont {Stockley}},\ }\href@noop {} {\bibfield  {journal} {\bibinfo  {journal} {Current opinion in virology}\ }\textbf {\bibinfo {volume} {31}},\ \bibinfo {pages} {74} (\bibinfo {year} {2018})}\BibitemShut {NoStop}%
\bibitem [{\citenamefont {Adhikari}\ \emph {et~al.}(2020)\citenamefont {Adhikari}, \citenamefont {Li}, \citenamefont {Shin}, \citenamefont {Steinmetz}, \citenamefont {Twarock}, \citenamefont {Podgornik},\ and\ \citenamefont {Ching}}]{adhikari2020intra}%
  \BibitemOpen
  \bibfield  {author} {\bibinfo {author} {\bibfnamefont {P.}~\bibnamefont {Adhikari}}, \bibinfo {author} {\bibfnamefont {N.}~\bibnamefont {Li}}, \bibinfo {author} {\bibfnamefont {M.}~\bibnamefont {Shin}}, \bibinfo {author} {\bibfnamefont {N.~F.}\ \bibnamefont {Steinmetz}}, \bibinfo {author} {\bibfnamefont {R.}~\bibnamefont {Twarock}}, \bibinfo {author} {\bibfnamefont {R.}~\bibnamefont {Podgornik}}, \ and\ \bibinfo {author} {\bibfnamefont {W.-Y.}\ \bibnamefont {Ching}},\ }\href@noop {} {\bibfield  {journal} {\bibinfo  {journal} {Physical Chemistry Chemical Physics}\ }\textbf {\bibinfo {volume} {22}},\ \bibinfo {pages} {18272} (\bibinfo {year} {2020})}\BibitemShut {NoStop}%
\bibitem [{\citenamefont {Borukhov}\ \emph {et~al.}(1999)\citenamefont {Borukhov}, \citenamefont {Andelman},\ and\ \citenamefont {Orland}}]{borukh}%
  \BibitemOpen
  \bibfield  {author} {\bibinfo {author} {\bibfnamefont {I.}~\bibnamefont {Borukhov}}, \bibinfo {author} {\bibfnamefont {D.}~\bibnamefont {Andelman}}, \ and\ \bibinfo {author} {\bibfnamefont {H.}~\bibnamefont {Orland}},\ }\href@noop {} {\bibfield  {journal} {\bibinfo  {journal} {J. Phys. Chem. B}\ }\textbf {\bibinfo {volume} {103}},\ \bibinfo {pages} {5042} (\bibinfo {year} {1999})}\BibitemShut {NoStop}%
\bibitem [{\citenamefont {Shafir}\ \emph {et~al.}(2003)\citenamefont {Shafir}, \citenamefont {Andelman},\ and\ \citenamefont {Netz}}]{shafir}%
  \BibitemOpen
  \bibfield  {author} {\bibinfo {author} {\bibfnamefont {A.}~\bibnamefont {Shafir}}, \bibinfo {author} {\bibfnamefont {D.}~\bibnamefont {Andelman}}, \ and\ \bibinfo {author} {\bibfnamefont {R.~R.}\ \bibnamefont {Netz}},\ }\href@noop {} {\bibfield  {journal} {\bibinfo  {journal} {J. Chem. Phys.}\ }\textbf {\bibinfo {volume} {119}},\ \bibinfo {pages} {2355} (\bibinfo {year} {2003})}\BibitemShut {NoStop}%
\bibitem [{\citenamefont {Siber}\ and\ \citenamefont {Podgornik}(2008)}]{siber_rudipod08}%
  \BibitemOpen
  \bibfield  {author} {\bibinfo {author} {\bibfnamefont {A.}~\bibnamefont {Siber}}\ and\ \bibinfo {author} {\bibfnamefont {R.}~\bibnamefont {Podgornik}},\ }\href@noop {} {\bibfield  {journal} {\bibinfo  {journal} {Phys. Rev. E}\ }\textbf {\bibinfo {volume} {78}},\ \bibinfo {pages} {051915} (\bibinfo {year} {2008})}\BibitemShut {NoStop}%
\bibitem [{\citenamefont {Podgornik}(1991)}]{rudi91}%
  \BibitemOpen
  \bibfield  {author} {\bibinfo {author} {\bibfnamefont {R.}~\bibnamefont {Podgornik}},\ }\href@noop {} {\bibfield  {journal} {\bibinfo  {journal} {J. Phys. Chem.}\ }\textbf {\bibinfo {volume} {95}},\ \bibinfo {pages} {5249} (\bibinfo {year} {1991})}\BibitemShut {NoStop}%
\bibitem [{\citenamefont {Andelman}(1995)}]{andelman}%
  \BibitemOpen
  \bibfield  {author} {\bibinfo {author} {\bibfnamefont {D.}~\bibnamefont {Andelman}},\ }\href@noop {} {\emph {\bibinfo {title} {Structure and Dynamics of Membranes Chap. 12}}},\ edited by\ \bibinfo {editor} {\bibfnamefont {E.~S.}\ \bibnamefont {R.~Lipowsky}},\ Vol.~\bibinfo {volume} {1B}\ (\bibinfo  {publisher} {Elsevier, Amsterdam},\ \bibinfo {address} {Handbook of Biological Physics},\ \bibinfo {year} {1995})\BibitemShut {NoStop}%
\bibitem [{\citenamefont {Budkov}\ and\ \citenamefont {Kalikin}(2023)}]{PhysRevE.107.024503}%
  \BibitemOpen
  \bibfield  {author} {\bibinfo {author} {\bibfnamefont {Y.~A.}\ \bibnamefont {Budkov}}\ and\ \bibinfo {author} {\bibfnamefont {N.~N.}\ \bibnamefont {Kalikin}},\ }\href@noop {} {\bibfield  {journal} {\bibinfo  {journal} {Phys. Rev. E}\ }\textbf {\bibinfo {volume} {107}},\ \bibinfo {pages} {024503} (\bibinfo {year} {2023})}\BibitemShut {NoStop}%
\bibitem [{\citenamefont {Muthukumar}(1989)}]{Muthudisorder}%
  \BibitemOpen
  \bibfield  {author} {\bibinfo {author} {\bibfnamefont {M.}~\bibnamefont {Muthukumar}},\ }\href@noop {} {\bibfield  {journal} {\bibinfo  {journal} {The Journal of Chemical Physics}\ }\textbf {\bibinfo {volume} {90}},\ \bibinfo {pages} {4594} (\bibinfo {year} {1989})}\BibitemShut {NoStop}%
\bibitem [{\citenamefont {Hribar-Lee}\ \emph {et~al.}(2011)\citenamefont {Hribar-Lee}, \citenamefont {Luk\v{s}i\v{c}},\ and\ \citenamefont {Vlachy}}]{C1PC90001C}%
  \BibitemOpen
  \bibfield  {author} {\bibinfo {author} {\bibfnamefont {B.}~\bibnamefont {Hribar-Lee}}, \bibinfo {author} {\bibfnamefont {M.}~\bibnamefont {Luk\v{s}i\v{c}}}, \ and\ \bibinfo {author} {\bibfnamefont {V.}~\bibnamefont {Vlachy}},\ }\href {\doibase 10.1039/C1PC90001C} {\bibfield  {journal} {\bibinfo  {journal} {Annu. Rep. Prog. Chem.{,} Sect. C: Phys. Chem.}\ }\textbf {\bibinfo {volume} {107}},\ \bibinfo {pages} {14} (\bibinfo {year} {2011})}\BibitemShut {NoStop}%
\bibitem [{\citenamefont {Bratko}\ and\ \citenamefont {Chakraborty}(1995)}]{bratko1995polyelectrolyte}%
  \BibitemOpen
  \bibfield  {author} {\bibinfo {author} {\bibfnamefont {D.}~\bibnamefont {Bratko}}\ and\ \bibinfo {author} {\bibfnamefont {A.}~\bibnamefont {Chakraborty}},\ }\href@noop {} {\bibfield  {journal} {\bibinfo  {journal} {Physical Review E}\ }\textbf {\bibinfo {volume} {51}},\ \bibinfo {pages} {5805} (\bibinfo {year} {1995})}\BibitemShut {NoStop}%
\bibitem [{\citenamefont {Binder}\ and\ \citenamefont {Young}(1986)}]{binder_young}%
  \BibitemOpen
  \bibfield  {author} {\bibinfo {author} {\bibfnamefont {K.}~\bibnamefont {Binder}}\ and\ \bibinfo {author} {\bibfnamefont {A.~P.}\ \bibnamefont {Young}},\ }\href@noop {} {\bibfield  {journal} {\bibinfo  {journal} {Rev. Mod. Phys.}\ }\textbf {\bibinfo {volume} {58}},\ \bibinfo {pages} {801} (\bibinfo {year} {1986})}\BibitemShut {NoStop}%
\bibitem [{\citenamefont {Parisi}(2023)}]{parisi2023}%
  \BibitemOpen
  \bibfield  {author} {\bibinfo {author} {\bibfnamefont {G.}~\bibnamefont {Parisi}},\ }\href@noop {} {\bibfield  {journal} {\bibinfo  {journal} {Rev. Mod. Phys.}\ }\textbf {\bibinfo {volume} {95}},\ \bibinfo {pages} {030501} (\bibinfo {year} {2023})}\BibitemShut {NoStop}%
\bibitem [{\citenamefont {Garel}\ \emph {et~al.}(1997)\citenamefont {Garel}, \citenamefont {Orland},\ and\ \citenamefont {Pitard}}]{g.o.p}%
  \BibitemOpen
  \bibfield  {author} {\bibinfo {author} {\bibfnamefont {T.}~\bibnamefont {Garel}}, \bibinfo {author} {\bibfnamefont {H.}~\bibnamefont {Orland}}, \ and\ \bibinfo {author} {\bibfnamefont {E.}~\bibnamefont {Pitard}},\ }\href@noop {} {\emph {\bibinfo {title} {Spin glasses and random fields}}},\ edited by\ \bibinfo {editor} {\bibfnamefont {A.}~\bibnamefont {Young}}\ (\bibinfo  {publisher} {World Scientific},\ \bibinfo {year} {1997})\BibitemShut {NoStop}%
\bibitem [{\citenamefont {Li}\ \emph {et~al.}(2018)\citenamefont {Li}, \citenamefont {Orland},\ and\ \citenamefont {Zandi}}]{Li_2018}%
  \BibitemOpen
  \bibfield  {author} {\bibinfo {author} {\bibfnamefont {S.}~\bibnamefont {Li}}, \bibinfo {author} {\bibfnamefont {H.}~\bibnamefont {Orland}}, \ and\ \bibinfo {author} {\bibfnamefont {R.}~\bibnamefont {Zandi}},\ }\href {\doibase 10.1088/1361-648X/aab0c6} {\bibfield  {journal} {\bibinfo  {journal} {Journal of Physics: Condensed Matter}\ }\textbf {\bibinfo {volume} {30}},\ \bibinfo {pages} {144002} (\bibinfo {year} {2018})}\BibitemShut {NoStop}%
\bibitem [{\citenamefont {Naji}\ \emph {et~al.}(2013)\citenamefont {Naji}, \citenamefont {Kandu\v{c}}, \citenamefont {Forsman},\ and\ \citenamefont {Podgornik}}]{Perspective}%
  \BibitemOpen
  \bibfield  {author} {\bibinfo {author} {\bibfnamefont {A.}~\bibnamefont {Naji}}, \bibinfo {author} {\bibfnamefont {M.}~\bibnamefont {Kandu\v{c}}}, \bibinfo {author} {\bibfnamefont {J.}~\bibnamefont {Forsman}}, \ and\ \bibinfo {author} {\bibfnamefont {R.}~\bibnamefont {Podgornik}},\ }\href@noop {} {\bibfield  {journal} {\bibinfo  {journal} {The Journal of Chemical Physics}\ }\textbf {\bibinfo {volume} {139}},\ \bibinfo {pages} {150901} (\bibinfo {year} {2013})}\BibitemShut {NoStop}%
\bibitem [{\citenamefont {Markovich}\ \emph {et~al.}(2021)\citenamefont {Markovich}, \citenamefont {Andelman},\ and\ \citenamefont {Podgornik}}]{markovich2021charged}%
  \BibitemOpen
  \bibfield  {author} {\bibinfo {author} {\bibfnamefont {T.}~\bibnamefont {Markovich}}, \bibinfo {author} {\bibfnamefont {D.}~\bibnamefont {Andelman}}, \ and\ \bibinfo {author} {\bibfnamefont {R.}~\bibnamefont {Podgornik}},\ }in\ \href@noop {} {\emph {\bibinfo {booktitle} {Handbook of lipid membranes}}}\ (\bibinfo  {publisher} {CRC Press},\ \bibinfo {year} {2021})\ pp.\ \bibinfo {pages} {99--128}\BibitemShut {NoStop}%
\bibitem [{\citenamefont {Podgornik}\ and\ \citenamefont {Li\v{c}er}(2006)}]{bridg2006}%
  \BibitemOpen
  \bibfield  {author} {\bibinfo {author} {\bibfnamefont {R.}~\bibnamefont {Podgornik}}\ and\ \bibinfo {author} {\bibfnamefont {M.}~\bibnamefont {Li\v{c}er}},\ }\href@noop {} {\bibfield  {journal} {\bibinfo  {journal} {Current Opinion in Colloid \& Interface Science}\ }\textbf {\bibinfo {volume} {11}},\ \bibinfo {pages} {273} (\bibinfo {year} {2006})}\BibitemShut {NoStop}%
\bibitem [{\citenamefont {{de Gennes}}(1987)}]{DEGENNES1987189}%
  \BibitemOpen
  \bibfield  {author} {\bibinfo {author} {\bibfnamefont {P.}~\bibnamefont {{de Gennes}}},\ }\href@noop {} {\bibfield  {journal} {\bibinfo  {journal} {Advances in Colloid and Interface Science}\ }\textbf {\bibinfo {volume} {27}},\ \bibinfo {pages} {189} (\bibinfo {year} {1987})}\BibitemShut {NoStop}%
\bibitem [{\citenamefont {Shiferaw}\ and\ \citenamefont {Goldschmidt}(2001)}]{shiferaw}%
  \BibitemOpen
  \bibfield  {author} {\bibinfo {author} {\bibfnamefont {Y.}~\bibnamefont {Shiferaw}}\ and\ \bibinfo {author} {\bibfnamefont {Y.~Y.}\ \bibnamefont {Goldschmidt}},\ }\href@noop {} {\bibfield  {journal} {\bibinfo  {journal} {Phys. Rev. E}\ }\textbf {\bibinfo {volume} {63}},\ \bibinfo {pages} {051803} (\bibinfo {year} {2001})}\BibitemShut {NoStop}%
\bibitem [{\citenamefont {Anderson}(1958)}]{anderson1958}%
  \BibitemOpen
  \bibfield  {author} {\bibinfo {author} {\bibfnamefont {P.~W.}\ \bibnamefont {Anderson}},\ }\href {\doibase 10.1103/PhysRev.109.1492} {\bibfield  {journal} {\bibinfo  {journal} {Phys. Rev.}\ }\textbf {\bibinfo {volume} {109}},\ \bibinfo {pages} {1492} (\bibinfo {year} {1958})}\BibitemShut {NoStop}%
\bibitem [{\citenamefont {Crisanti}\ \emph {et~al.}(2012)\citenamefont {Crisanti}, \citenamefont {Paladin},\ and\ \citenamefont {Vulpiani}}]{crisanti_matrix}%
  \BibitemOpen
  \bibfield  {author} {\bibinfo {author} {\bibfnamefont {A.}~\bibnamefont {Crisanti}}, \bibinfo {author} {\bibfnamefont {G.}~\bibnamefont {Paladin}}, \ and\ \bibinfo {author} {\bibfnamefont {A.}~\bibnamefont {Vulpiani}},\ }\href {https://books.google.am/books?id=ogLtCAAAQBAJ} {\emph {\bibinfo {title} {Products of Random Matrices: in Statistical Physics}}},\ Springer Series in Solid-State Sciences\ (\bibinfo  {publisher} {Springer Berlin Heidelberg},\ \bibinfo {year} {2012})\BibitemShut {NoStop}%
\end{thebibliography}%

\newpage

\appendix

\renewcommand{\theequation}{A.\arabic{equation}}
\setcounter{equation}{0}

{\bf Appendices}

\section{The free energy functional}\label{appendix:FEF}

The quenched free energy (Eq. \ref{freen_self}) can be estimated using the replica trick \cite{binder_young}
\begin{equation}
-\beta{\cal F}=\lim_{n\rightarrow 0}\frac{\langle{Z\{\xi\}^n}\rangle_{{\cal P}}-1}{n},
\end{equation}
where $\beta=(k_{B}T)^{-1}$. 
We therefore need to calculate the $n$-replica partition function as
\begin{eqnarray}\label{Z^n_app}
&&\langle{Z\{\xi\}^n}\rangle_{{\cal P}} =\notag \\ 
&& \int \,{\cal D}{\bf r}
e^{-\frac{3}{2\ell^2}\sum_{a=1}^n\int_0^N \,d\tau(\partial_{\tau}{\bf r}^a(\tau))^2- 
\beta\sum_{a=1}^n V\{{\bf r}^a\}}\times \notag \\ 
&& \int \,{\cal D}\xi{\cal P}\{\xi\}e^{-\frac{\beta v_0}{2}\int_0^N \,d\tau\int_0^N \,d\tau'\xi_{\tau}\xi_{\tau'}\sum_{a=1}^n
\delta({\bf r}^a(\tau)-{\bf r}^a(\tau'))}.~~~~~
\end{eqnarray}
The average over the distribution function (Eq. \ref{distribfunc}) in the equation (Eq. \ref{Z^n_app}) is transformed as
\begin{eqnarray}\label{transf} 
\int \,{\cal D}\xi{\cal P}\{\xi\}e^{-\frac{\beta v_0}{2}\int_0^N \,d\tau\int_0^N \,d\tau'\xi_{\tau}\xi_{\tau'}\sum_{a=1}^n
\delta({\bf r}^a(\tau)-{\bf r}^a(\tau'))}=\notag \\ 
\int \,{\cal D}\xi e^{-\frac{1}{2\xi^2}\int_0^N \,d\tau \xi_{\tau}^2}
e^{-\frac{\beta v_0}{2}\int \,d^3{\bf x}\sum_{a=1}^n m_a({\bf x})^2},~~~~
\end{eqnarray} 
where the field $m_a({\bf x})=\int_0^N \,d\tau\xi_{\tau}\delta({\bf x}-{\bf r}^a(\tau))$, and $v_0$ is the interaction strength parameter. The r.h.s. of the equation (Eq. \ref{transf}) is linearized using the Hubbard-Stratonovich transformation as
\begin{eqnarray}\label{r.h.s.} 
e^{-\frac{\beta v_0}{2}\int \,d^3{\bf x}\sum_{a=1}^n m_a({\bf x})^2}=e^{-\frac{nV}{2}\ln(2\pi\beta v_0)}\notag \\ 
\int \,{\cal D}{\Pi}e^{-\frac{1}{2\beta v_0}\sum_{a}\int \,d^3{\bf x}\Pi_{a}({\bf x})^2+\imath\sum_{a}\int \,d^3{\bf x}\Pi_{a}({\bf x})m_{a}({\bf x})}.
\end{eqnarray}
Insertion of (Eqs.~\ref{transf},\ref{r.h.s.}) into the (Eq.~\ref{Z^n_app}) upon averaging over the $\{\xi\}$ variables 
yields the $n$-replica partition function as 
\begin{align}\label{Z^n_2}
\langle{Z\{\xi\}^n}\rangle_{{\cal P}}\propto e^{-\frac{nV}{2\tilde{v}}\ln(2\pi\beta v_0)}
\int \,{\cal D}{\Pi}e^{-\frac{1}{2\beta v_0}\sum_{a}\int \,d^3{\bf x}\Pi_{a}({\bf x})^2}\times \notag \\ 
\int \,{\cal D}{\bf r}
e^{-\frac{3}{2\ell^2}\sum_{a=1}^n\int_0^N \,d\tau(\partial_{\tau}{\bf r}^a(\tau))^2-
\beta\sum_{a=1}^nV\{{\bf r}^a\}}\times \notag \\ 
e^{-\frac{\xi^2}{2}\sum_{a,b}\int_0^N \,d\tau\Pi_{a}({\bf r}^a(\tau))\Pi_{b}({\bf r}^b(\tau))}.
\end{align}
For further consideration it is useful to introduce two order parameters, {\sl viz.},  the inter-replica overlap $q_{ab}({\bf x},{\bf x}')$ , with $a<b$, and 
the density of $a$-th replica monomers $\rho_a({\bf x})$ as
\begin{eqnarray}\label{q_and_rho}
q_{ab}({\bf x},{\bf x}')=\int_0^N \,d\tau\delta({\bf x}-{\bf r}^a(\tau))\delta({\bf x}'-{\bf r}^b(\tau))\notag \\ 
\rho_a({\bf x})=\int_0^N \,d\tau\delta({\bf x}-{\bf r}^a(\tau)).
\end{eqnarray}
Using these order parameters (Eq. \ref{q_and_rho}), the $n$-replica partition function (Eq. \ref{Z^n_2}) is transformed as
\begin{align}\label{Z^n_3}
\langle{Z\{\xi\}^n}\rangle_{{\cal P}}\propto 
\int\!\!{\cal D}{\varphi}\!\!\int\!\!{\cal D}{c^{\pm}}\!\!\int\!\!{\cal D}{\Pi}~
e^{-\frac{1}{2\beta v_0}\sum_{a}\int \,d^3{\bf x}\Pi_{a}({\bf x})^2}\times \notag \\ 
\int \,{\cal D}{\bf r}
e^{-\frac{3}{2\ell^2}\sum_{a=1}^n\int_0^N \,d\tau(\partial_{\tau}{\bf r}^a(\tau))^2}
\times \notag \\ 
e^{-\beta\sum_{a=1}^n W_{el}(\rho_a,\varphi_a,c_a^{\pm})-\frac{\xi^2}{2}\sum_{a}\int \,d^3{\bf x}\Pi_{a}({\bf x})^2\rho_a({\bf x})}\times \notag \\ 
e^{-\xi^2\sum_{a<b}\int \,d^3{\bf x}\int \,d^3{\bf x}'\Pi_{a}({\bf x})\Pi_{b}({\bf x}')q_{ab}({\bf x},{\bf x}')},~~~~~
\end{align} 
where the electrostatic part of the free energy $W_{el}(\rho_a,\varphi_a,c_a^{\pm})$ is written as a sum of electrostatic energies of polyelectrolyte segments, fixed charges residing on the surface, confining polyelectrolyte and charges of salt ions, as well as the entropies of the salt ions. As shown in \cite{borukh,shafir,siber_rudipod08},
\begin{align}
W_{el}(\rho_a,\varphi_a,c_a^{\pm})=\int\,d^3{\bf x}\bigg\{-\frac{\epsilon\epsilon_0}{2}(\nabla\varphi_a({\bf x}))^2+\notag \\ 
\varphi_a({\bf x})\bigg[ec_a^{+}({\bf x})-ec_a^{-}({\bf x})
-pe\rho_a({\bf x})+\rho_{surf}({\bf x})\bigg]+\notag \\ 
\sum_{i=\pm}\bigg[k_BT(c_a^i({\bf x})\ln c_a^i({\bf x})-\notag \\ 
c_a^i({\bf x})-(c_0^i\ln c_0^i-c_0^i))-\mu^i(c_a^i({\bf x})-c_0^i)\bigg]\bigg\}.
\end{align}
Here $c^{\pm}$ are the concentrations of $\pm$ monovalent salt ions, with $c_0^{\pm}$ being their bulk concentrations, and $\mu^{\pm}$ their chemical potentials, $\varepsilon\varepsilon_0$ is the permittivity of water, and $\rho_{surf}$ is the charge distribution over the bounding surfaces, confining polyelectrolyte.

Introducing the Lagrange multipliers ${\hat q}_{ab}({\bf x},{\bf x}')$ and the ${\hat \rho}_a({\bf x})$ conjugated with the 
order parameters (\ref{q_and_rho}), the $n$-replica partition function reads \cite{g.o.p}
\begin{align}\label{Z^n_ord_params_app}
\langle{Z\{\xi\}^n}\rangle_{{\cal P}}\propto e^{-\frac{nV}{2\tilde{v}}\ln(2\pi\beta v_0)}
\int \,{\cal D}{\Pi}\,{\cal D}{\rho}\,{\cal D}{{\hat\rho}}\,{\cal D}{q}\,{\cal D}{\hat q} \notag \\
e^{G(\Pi,\rho,{\hat \rho},q,{\hat q})+\ln\zeta({\hat \rho},{\hat q})},
\end{align}
where
\begin{align}
G(\Pi,\rho,{\hat \rho},q,{\hat q},\varphi,c^{\pm})=-\frac{1}{2\beta v_0}\sum_{a}\int \,d^3{\bf x}\Pi_{a}({\bf x})^2-\notag \\ 
-\beta \sum_{a=1}^n W_{el}(\rho_a,\varphi_a,c_a^{\pm})
-\frac{\xi^2}{2}\sum_{a}\int \,d^3{\bf x}\Pi_{a}({\bf x})^2\rho_a({\bf x})-\notag \\ 
-\xi^2\sum_{a<b}\int \,d^3{\bf x}\int \,d^3{\bf x}'\Pi_{a}({\bf x})\Pi_{b}({\bf x}')q_{ab}({\bf x},{\bf x}')+\notag \\ 
+\imath\sum_{a}\int \,d^3{\bf x}\rho_a({\bf x}){\hat \rho}_a({\bf x})+ 
\imath\sum_{a<b}\int \,d^3{\bf x}\int \,d^3{\bf x}'q_{ab}({\bf x},{\bf x}'){\hat q}_{ab}({\bf x},{\bf x}')
\end{align}
and
\begin{eqnarray}
\ln\zeta({\hat \rho},{\hat q})=-N\min_{\psi}\bigg\{\int \,d^{3n}{\bf x}\psi({\bf x}_1,...,{\bf x}_n)\times\notag \\ 
\bigg(-\frac{\ell^2}{6}\sum_{a}\nabla_{a}^2
+\imath\sum_{a}{\hat \rho}_a({\bf x}_a)+\imath\sum_{a<b}{\hat q}_{ab}({\bf x}_a,{\bf x}_b)\bigg)\times \notag \\ 
\psi({\bf x}_1,...,{\bf x}_n)
-\mathcal{E}_0\bigg(\int \,d^{3n}{\bf x}\psi({\bf x}_1,...,{\bf x}_n)^2-1\bigg)\bigg\}.
\end{eqnarray}
Here $\mathcal{E}_0$ is the ground state energy and $\psi({\bf x}_1,...,{\bf x}_n)$ is the corresponding eigenfunction of the 
``quantum-like" Hamiltonian ${\cal {\hat H}}_n=-\frac{\ell^2}{6}\sum_{a}\nabla_{a}^2+\imath\sum_{a}{\hat \rho}_a({\bf x}_a)+\imath\sum_{a<b}{\hat q}_{ab}({\bf x}_a,{\bf x}_b)$. 
Since $v_0>0$, we have no trend of segregation between different kinds of monomers and we expect large fluctuations of the fields $\{\Pi\}$. Thus, we can consider the saddle-point approximation in the equation (Eq. \ref{Z^n_ord_params_app}) over the fields $\{\rho,{\hat \rho},q,{\hat q}\}$ only and to integrate over the fields $\{\Pi\}$. Let us introduce $g(\rho,{\hat \rho},q,{\hat q},\varphi,c^{\pm})=\ln\int \,{\cal D}{\Pi}~e^{G(\Pi,\rho,{\hat \rho},q,{\hat q})}$, where 
\begin{eqnarray}
\int \,{\cal D}{\Pi}~e^{G(\Pi,\rho,{\hat \rho},q,{\hat q})}=
e^{-\beta\sum_{a=1}^n W_{el}(\rho_a,\varphi_a,c_a^{\pm})}\times \notag \\ 
e^{\imath\sum_{a}\int \,d^3{\bf x}\rho_a({\bf x}){\hat \rho}_a({\bf x})+
\imath\sum_{a<b}\int \,d^3{\bf x}\int \,d^3{\bf x}'q_{ab}({\bf x},{\bf x}'){\hat q}_{ab}({\bf x},{\bf x}')}\times
\notag \\ 
\int \,{\cal D}{\Pi}e^{-\frac{1}{2}\sum_{a,b}\int \,d^3{\bf x}\int \,d^3{\bf x}'P_{ab}({\bf x},{\bf x}')\Pi_{a}({\bf x})\Pi_{b}({\bf x}')},
\end{eqnarray}
and $P_{ab}({\bf x},{\bf x}')=\delta_{ab}\delta({\bf x}-{\bf x}')[\frac{1}{\beta v_0}+\xi^2\rho_a({\bf x})]+
\xi^2(1-\delta_{ab})q_{ab}({\bf x},{\bf x}')$. Thus, $g(\rho,{\hat \rho},q,{\hat q})$ can be obtained as
\begin{eqnarray}\label{g}
g(\rho,{\hat \rho},q,{\hat q},\varphi,c^{\pm})=-\frac{1}{2}\ln\det\frac{{\hat P}}{2\pi}-\notag \\ 
\beta\sum_{a=1}^n W_{el}(\rho_a,\varphi_a,c_a^{\pm})
+\imath\sum_{a}\int \,d^3{\bf x}\rho_a({\bf x}){\hat \rho}_a({\bf x})+\notag \\ 
\imath\sum_{a<b}\int \,d^3{\bf x}\int \,d^3{\bf x}'q_{ab}({\bf x},{\bf x}'){\hat q}_{ab}({\bf x},{\bf x}').
\end{eqnarray}
Let us consider the term $\ln\det$ in the equation (Eq. \ref{g}). These calculations may be carried out by employing properties of block-matrices and the operator algebra defined over the Hilbert space
$\{|{\bf x}\rangle\}$. We shall use the compact notation by defining the operators
$\langle{\bf x}|{\hat A}_{ab}|{\bf x}'\rangle =\delta_{ab}\delta({\bf x}-{\bf x}')[\frac{1}{\beta v_0}+\xi^2\rho_a({\bf x})]$ as the $n\times n$ operator matrix
and $\langle{\bf x}|{\hat P}_{ab}|{\bf x'}\rangle = P_{ab}({\bf x},{\bf x'})$ as  
an element of the $n\times n$ operator matrix
\begin{equation}
{\hat P}={\hat A}+\xi^2{\hat B},
\end{equation}
where $\langle{\bf x}|{\hat B}_{ab}|{\bf x'}\rangle=
(1-\delta_{ab})q_{ab}({\bf x},{\bf x}')$ is the $n\times n$ operator matrix. Therefore
\begin{align}\label{lndet}
\ln\det{\hat P}={\rm Tr}\ln{\hat P}={\rm Tr}\ln\bigg\{{\hat A}+
\xi^2{\hat B}\bigg\}={\rm Tr}\ln{\hat A}+\notag \\ 
{\rm Tr}\ln\bigg\{{\hat 1}_n\otimes{\hat 1}+\xi^2{\hat A}^{-1}{\hat B}\bigg\},
\end{align}
where $({\hat 1}_n)_{ab}=\delta_{ab}$ and $\langle{\bf x}|{\hat 1}|{\bf x}'\rangle=\delta({\bf x}-{\bf x}')$. Calculation of the 
first term on the r.h.s. of the equation (Eq. \ref{lndet}) is straightforward
\begin{align}\label{TrlnA}
	{\rm Tr}\ln{\hat A}= \sum_{k=1}^{\infty}\frac{(-1)^{k+1}}{k}\sum_{a}\int\,d^3{\bf x}\langle{\bf x}|({\hat A} - {\hat 1})^k_{aa}|{\bf x}\rangle = \notag \\ 
	= \sum_{k=1}^{\infty}\frac{(-1)^{k+1}}{k}\sum_{a}\int\,d^3{\bf x} \delta({\bf x}-{\bf x}) [\frac{1}{\beta v_0}+\xi^2\rho_a({\bf x})-1]^k = \notag \\ 
	=  \sum_{a} \int\,d^3{\bf x}\delta({\bf x}-{\bf x})\ln[\frac{1}{\beta v_0}+\xi^2\rho_a({\bf x})] \approx \notag \\ 
	\approx  \frac{1}{\tilde{v}}\sum_{a}\int \,d^3{\bf x}\ln\bigg(\frac{1}{\beta v_0}+\xi^2\rho_{a}({\bf x})\bigg),
\end{align}

using $\delta(0) \approx \frac{1}{\tilde{v}}$, where $\tilde{v}$ is volume of the monomer.

\renewcommand{\theequation}{B.\arabic{equation}}
\setcounter{equation}{0}

\section{Replica symmetric solution}\label{appendix:RSS}

In the replica-symmetric case we will consider only the Hartree-like replica-symmetric wave functions
\begin{equation}\label{Hartree}
\psi({\bf x}_1,...,{\bf x}_n) = \prod_{a=1}^n \psi({\bf x}_a),
\end{equation}	
where $\psi({\bf x}_a)$ is the one-replica wave function.
In the replica symmetric regime the expression (Eq. \ref{cal_F}) may be written (using (Eq. \ref{TrlnA})) as
\begin{align}\label{cal_F_rss_s}
-\beta {\mathcal F} = \lim_{n\rightarrow 0}\frac{1}{n}\bigg[-\frac{n}{2\tilde{v}}\int \,d^3{\bf x}\ln\bigg(1+\frac{\rho_{a}({\bf x})}{\mu}\bigg)- \frac{1}{2}{\rm Tr}\ln\bigl\{I + \hat C\bigr\}- \notag\\ 
-\beta n W_{el} + i n \langle\rho|\hat\rho\rangle 
+i\frac{n(n-1)}{2}\langle q|\hat q\rangle+\ln\zeta\bigg],
\end{align}
where $\hat C = \xi^2\hat A^{-1}\hat B$ and $\mu = (\beta v_0 \xi^2)^{-1}$. With the replica symmetry we have $\hat{C} = \hat{T} \otimes \hat{C'}$ where $\hat{T}_{ab} = 1-\delta_{ab}$ and $\langle x|\hat{C'}|x'\rangle = \frac{q(x,x')}{\mu+\rho(x)}$. Taking the limit we get the reduced free energy
\begin{align}\label{cal_F_rss_s1}
f = -\beta {\mathcal F} =-\frac{1}{2\tilde{v}}\int \,d^3{\bf x}\ln\bigg(1+\frac{\rho_{a}({\bf x})}{\mu}\bigg)- \frac{\Xi}{2}-\notag\\ -\beta W_{el} + i \langle\rho|\hat\rho\rangle 
-\frac{i}{2}\langle q|\hat q\rangle+\ln\zeta,
\end{align}
where we defined
\begin{equation}\label{Xi_dfn}
	\Xi\equiv\lim_{n\to0} \frac{1}{n}{\rm Tr}\ln\bigl\{I + \hat C\bigr\} =
\lim_{n\to0} \frac{1}{n}\sum_{k=1}^{\infty}\frac{(-1)^{k+1}}{k}{\rm Tr}(\hat{C}^{k}). 
\end{equation} Thus, using ${\rm Tr}(\hat{T} \otimes \hat{C'}) = {\rm Tr}(\hat{T}){\rm Tr}(\hat{C'})$, the terms of the series are written
	\begin{align}{\rm Tr}(\hat C)^k = T_k \int d^{3k}x \frac{q(x^{(1)},x^{(2)})q(x^{(2)},x^{(3)})...q(x^{(k)},x^{(1)})}{(\mu + \rho(x^{(1)}))(\mu + \rho(x^{(2)}))..(\mu + \rho(x^{(k)}))}\notag\\
	T_k = \sum_{a_1,..,a_k}\bigg[\prod_{i=2}^{k}\biggl\{(1-\delta_{a_i,a_{i-1}})\biggr\}(1-\delta_{a_k,a_1})\bigg]
	\end{align}

Where $T_k = {\rm Tr}(\hat{T}^k)$. To calculate $T_k$ first notice the following recursive equation
\begin{equation}
T_k = T_{k-1}(n-2)+T_{k-2}(n-1)
\end{equation}
with the initial conditions $T_2 = n(n-1)$ and $T_3 = (n-2)T_2$. 
Let us define a generating function $P(x) = \sum_{i=0}^{\infty}x^iT_{i+2}$. Using 
\begin{eqnarray}
	x(n-2)P(x) = \sum_{i=1}^{\infty}x^iT_{i+1}(n-2) \\ \nonumber
	x^2(n-1)P(x) = \sum_{i=2}^{\infty}x^iT_{i}(n-1)
\end{eqnarray}
we can furthermore write
\begin{equation}
P(x)\bigg(1-(n-2)x-(n-1)x^2\bigg) = T_2 = n(n-1)
\end{equation}
and
\begin{equation}
P(x) = \sum_{i=0}^{\infty}x^i\bigg((n-1)^{i+2}-(-1)^{i+1}(n-1)\bigg)= \sum_{i=0}^{\infty}x^iT_{i+2}.
\end{equation}
Thus, we can take the limit
\begin{eqnarray}\label{tklimn0}
	\lim_{n\to0} \frac{T_k}{n} = \lim_{n\to0} \frac{(n-1)^{k}-(-1)^{k-1}(n-1)}{n} = ~~~~~~~~~~\nonumber\\ 
	= \lim_{n\to0} \bigg(k(n-1)^{k-1}-(-1)^{k-1}\bigg) = (-1)^{k-1}(k-1) ~~~~~~
\end{eqnarray}
Substituting (Eq. \ref{tklimn0}) into $\Xi$ defined by (Eq. \ref{Xi_dfn}) we get
\begin{equation}
\Xi=\sum_{k=2}^\infty \frac{k-1}{k} \int d^{3k}x \frac{q({\bf x}^{(1)},{\bf x}^{(2)})q({\bf x}^{(2)},{\bf x}^{(3)})...q({\bf x}^{(k)},{\bf x}^{(1)})}{(\mu + \rho({\bf x}^{(1)}))(\mu + \rho({\bf x}^{(2)}))..(\mu + \rho({\bf x}^{(k)}))}
\end{equation}

The saddle - point equations can be written in the following form 
	\begin{equation}\label{rho}
		0=\frac{\delta f}{\delta\hat{\rho}({\bf r})}\Rightarrow \rho({\bf r})=N\psi({\bf r})^2
	\end{equation}
	\begin{equation}\label{q}
		0=\frac{\delta f}{\delta\hat{q}({\bf r},{\bf r}')}\Rightarrow q({\bf r},{\bf r}') = N\psi({\bf r})^2\psi({\bf r}')^2
	\end{equation}
	\begin{equation}\label{rho_hat}
		0=\frac{\delta f}{\delta\rho({\bf r})}\Rightarrow \mathrm{i}\hat{\rho}({\bf r})=-\beta pe\varphi({\bf r})+\frac{1}{2\tilde{v}}\frac{1}{\mu+\rho(\bf r)}+\frac{1}{2}\frac{\delta \Xi}{\delta \rho}
	\end{equation}
	\begin{equation}\label{q_hat}
		0=\frac{\delta f}{\delta q({\bf r},{\bf r}')}\Rightarrow \mathrm{i}\hat{q}({\bf r},{\bf r}')=-\frac{\delta \Xi}{\delta q}
	\end{equation}
	\begin{equation}\label{varphi}
		0=\frac{\delta f}{\delta\varphi({\bf r})} \Rightarrow \varepsilon\varepsilon_0\nabla^2\varphi({\bf r})=-(ec^{+}-ec^{-}-pe\rho({\bf r})+\rho_s({\bf r})) 
	\end{equation}
	\begin{equation}\label{c}
		0=\frac{\delta f}{\delta c^{\pm}({\bf r})} \Rightarrow c^{\pm}({\bf r})=c_0\exp({\mp\beta e\varphi({\bf r})}) 
	\end{equation}
	\begin{align}\label{d_psi}
		0=\frac{\delta f}{\delta\psi({\bf r})}=-N \frac{\delta}{\delta\psi({\bf r})}\bigg\{\frac{\ell^2}{6}\langle\nabla\psi|\nabla\psi\rangle + \notag \\
  \mathrm{i}\langle\psi^2|\hat{\rho}\rangle- -\frac{\mathrm{i}}{2}\langle\psi^2|\hat{q}|\psi^2\rangle - \mathcal{E}_0(\langle\psi|\psi\rangle - 1)\bigg\} =\notag \\ 
  -N\bigg\{ -\frac{\ell^2}{3}\nabla^2 \psi({\bf r}) + 2\mathrm{i}\psi({\bf r})\hat{\rho}({\bf r}) -
  \mathrm{i}\psi({\bf r})\int\,d^3{\bf x}\hat{q}({\bf r},{\bf x})\psi({\bf x})^2 - \notag \\ 
  \mathrm{i}\psi({\bf r})\int\,d^3{\bf x}\hat{q}({\bf x},{\bf r})\psi({\bf x})^2 - 2\mathcal{E}_0\psi({\bf r})\bigg\}
	\end{align}
where
	\begin{equation}\begin{split}&\frac{\delta \Xi}{\delta \rho({\bf x})}=-\sum_{k=2}^\infty \frac{k-1}{(\mu + \rho({\bf x}))} \\ &\int d^{3k-3}{\bf x} \frac{q({\bf x},{\bf x}^{(1)})q({\bf x}^{(1)},{\bf x}^{(2)})...q({\bf x}^{(k-1)},{\bf x})}{(\mu + \rho({\bf x}))(\mu + \rho({\bf x}^{(1)}))..(\mu + \rho({\bf x}^{(k-1)}))}\end{split}\end{equation}
	\begin{equation}\begin{split}&\frac{\delta \Xi}{\delta q({\bf x},{\bf x}')}=\sum_{k=2}^\infty \frac{k-1}{(\mu + \rho({\bf x}))} \\ &\int d^{3k-6}{\bf x} \frac{q({\bf x}',{\bf x}^{(1)})q({\bf x}^{(1)},{\bf x}^{(2)})...q({\bf x}^{(k-2)},{\bf x})}{(\mu + \rho({\bf x}'))(\mu + \rho({\bf x}^{(1)}))..(\mu + \rho({\bf x}^{(k-2)}))}\end{split}
\end{equation}
By further defining $$\kappa\equiv\int d^3x \frac{N\psi({\bf x})^4}{\mu + N\psi({\bf x})^2}$$and using (Eq. \ref{rho}, \ref{q}) we get
\begin{equation}\label{Xi_by_rho}
\frac{\delta \Xi}{\delta \rho({\bf r})}=-\frac{\kappa}{(1-\kappa)^2}\frac{N\psi({\bf r})^4 }{(\mu + N\psi({\bf r})^2)^2}
\end{equation}
\begin{equation}\label{Xi_by_q}
\frac{\delta \Xi}{\delta q({\bf r},{\bf r}')}=\frac{1}{(1-\kappa)^2}\frac{N\psi({\bf r})^2\psi({\bf r}')^2 }{(\mu + N\psi({\bf r})^2)(\mu + N\psi({\bf r}')^2)}
\end{equation}
Substituting equations (Eqs. \ref{rho_hat},\ref{q_hat}) into the equation (Eq. \ref{d_psi}) we finally obtain
\begin{align}
		\frac{\ell^2}{6}\nabla^2 \psi({\bf r})=\psi({\bf r})\bigg[-\beta pe\varphi({\bf r})+\frac{1}{2\tilde{v}}\frac{1}{\mu+\rho(\bf r)}+\frac{1}{2}\frac{\delta \Xi}{\delta \rho}\bigg] +\notag \\
		+\psi({\bf r})\int\,d^3{\bf x}\frac{1}{2}\frac{\delta \Xi}{\delta q({\bf r},{\bf x})}\psi({\bf x})^2+\notag \\
  \psi({\bf r})\int\,d^3{\bf x}\frac{1}{2}\frac{\delta \Xi}{\delta q({\bf x},{\bf r})}\psi({\bf x})^2 - 
		-\mathcal{E}_0\psi({\bf r})
\end{align}
Together with equations (Eqs. \ref{Xi_by_rho},\ref{Xi_by_q}) the latter equation is transformed into
	\begin{align}\label{edwards_eqn}
		\frac{\ell^2}{6}\nabla^2 \psi({\bf r})=\psi({\bf r})\bigg[-\beta pe\varphi({\bf r}) - \mathcal{E}_0 +\frac{1}{2\tilde{v}}\frac{1}{\mu+N\psi({\bf r})^2} - \notag \\- \frac{\kappa}{(1-\kappa)^2}\frac{N\psi({\bf r})^4 }{2(\mu + N\psi({\bf r})^2)^2} + \notag \\ 
  \int\,d^3{\bf x}\frac{1}{(1-\kappa)^2}\frac{N\psi({\bf r})^2\psi({\bf x})^2 }{(\mu + N\psi({\bf r})^2)(\mu + N\psi({\bf x})^2}\psi({\bf x})^2)\bigg].
	\end{align}
 
\end{document}